\documentclass[12pt,draftcls,onecolumn]{IEEEtran}
\makeatletter
\def\ps@headings{%
\def\@oddhead{\mbox{}\scriptsize\rightmark \hfil \thepage}%
\def\@evenhead{\scriptsize\thepage \hfil \leftmark\mbox{}}%
\def\@oddfoot{}%
\def\@evenfoot{}}
\makeatother
\usepackage{nonfloat}
\usepackage{subfig}
\usepackage{multirow}
\usepackage{graphicx}
\usepackage{amssymb}
\usepackage{epstopdf}
\usepackage{amsmath}
\usepackage{amsthm}
\newtheorem{mydef}{Definition}
\newtheorem{lem}{Lemma}
\newtheorem{thm}{Theorem}
\newtheorem{cor}{Corollary}

\begin{document}
\title{An Upper Bound on the Convergence Time for Quantized Consensus}

%
%\numberofauthors{2}
%\author{
%\alignauthor
%Shang Shang\\
%       \affaddr{Department of Electrical Engineering}\\
%       \affaddr{Princeton University}\\
%       \affaddr{Princeton, NJ, 08540, U.S.A.}\\
%       \email{sshang@princeton.edu}
%% 2nd. author
%\alignauthor
%Paul W. Cuff\\
%	\affaddr{Department of Electrical Engineering}\\
%       \affaddr{Princeton University}\\
%       \affaddr{Princeton, NJ, 08540, U.S.A.}\\
%       \email{cuff@princeton.edu}
%% 3rd. author
%\and
%\alignauthor 
%Sanjeev R. Kulkarni\\
%       \affaddr{Department of Electrical Engineering}\\
%       \affaddr{Princeton University}\\
%       \affaddr{Princeton, NJ, 08540, U.S.A.}\\
%       \email{kulkarni@princeton.edu}
%% use '\and' if you need 'another row' of author names
%% 4th. author
%\alignauthor 
%Pan Hui\\
%       \affaddr{Deutsche Telekom Laboratories}\\
%       \affaddr{Ernst-Reuter-Platz 7}\\
%       \affaddr{10587 Berlin, Germany}\\
%       \email{pan.hui@telekom.de}
%       }
       
       \author{
\IEEEauthorblockN{Shang Shang\IEEEauthorrefmark{1},  Paul W. Cuff\IEEEauthorrefmark{1}, Pan
Hui\IEEEauthorrefmark{2}
and Sanjeev R. Kulkarni\IEEEauthorrefmark{1} }\\ \IEEEauthorblockA{\IEEEauthorrefmark{1}Department of Electrical Engineering, 
Princeton University \\ Princeton NJ, 08540, U.S.A.
} \\ \IEEEauthorblockA{\IEEEauthorrefmark{2}Deutsche Telekom
Laboratories, Ernst-Reuter-Platz 7, 10587 Berlin, Germany\\
\IEEEauthorrefmark{1}\{sshang, cuff, kulkarni\}@princeton.edu, \IEEEauthorrefmark{2}pan.hui@telekom.de}
}
       
\maketitle
\begin{abstract}
We analyze a class of distributed quantized consensus algorithms for arbitrary networks. In the initial setting, each node in the network has an integer value. Nodes exchange their current estimate of the mean value in the network, and then update their estimation by communicating with their neighbors in a limited capacity channel in an asynchronous clock setting. Eventually, all nodes reach consensus with quantized precision. We start the analysis with a special case of a distributed binary voting algorithm, then proceed to the expected convergence time for the general quantized consensus algorithm proposed by Kashyap et al. We use the theory of electric networks, random walks, and couplings of Markov chains to derive an $O(N^3\log N)$ upper bound for the expected convergence time on an arbitrary graph of size $N$, improving on the state of art bound of $O(N^4\log N)$ for binary consensus and $O(N^5)$ for quantized consensus algorithms. Our result is not dependent on graph topology. Simulations on special graphs such as star networks, line graphs, lollipop graphs, and Erd\"os-R\'enyi random graphs are performed to validate the analysis. 

This work has applications to load balancing, coordination of autonomous agents, estimation and detection, decision-making networks, peer-to-peer systems, etc.   
\end{abstract}
\section{Introduction}		
\label{sec:intro}

Over the past decade, the problem of quantized consensus has received  significant attention \cite{Boyd}\cite{Zhu}\cite{Benezit}\cite{Kashyap}\cite{Cai}. It models averaging in a network with a limited capacity channel \cite{Kashyap}. Distributed algorithms are attractive due to their flexibility, simple deployment and the lack of central control. This problem is of interest in the context of coordination of autonomous agents, estimation, distributed data fusion on sensor networks, peer-to-peer systems, etc. \cite{Boyd}\cite{Drief}. It is especially relevant to remote and extreme environments where communication and computation are limited, for example, in a decision-making sensor network \cite{Du}.
% as shown in Fig. \ref{sensor}.  

This work is motivated by a class of quantized consensus algorithms in \cite{Benezit} and in \cite{Kashyap}: nodes randomly and asynchronously update local estimate and exchange information.  In \cite{Benezit}, the author proposed a binary voting algorithm, where all the nodes in the network vote ``\emph{yes}" or ``\emph{no}".  The algorithm reaches consensus on the initial majority opinion almost surely. However, the authors did not bound the convergence time. In \cite{Mossel}, the authors studied the convergence speed in the special case of regular graphs for a similar distributed binary consensus algorithm. Draief and Vojnovic \cite{Drief} derived an expected convergence time bound depending on the second largest eigenvalue of a doubly stochastic matrix characterizing the algorithm and voting margin, yet no specific bound is provided for an arbitrary graph. An $O(N^4\log N)$ bound is given in \cite{Shang} on the binary voting convergence speed, where $N$ is the number of nodes in the network. A more general distributed quantized integer averaging algorithm was proposed in \cite{Kashyap}. Unlike the distributed algorithm in \cite{Nedic}, where the sum of values in the network is not preserved, Kashyap et al. proposed an algorithm guaranteeing convergence with limited communication, more specifically,  only involving quantization levels. This is a desired property in a large-scale network where memory is limited, communication between nodes is expensive and no central control is available to the network. Also, this distributed algorithm is designed in a privacy-preserving manner: during the process, the local estimation on the average value is exchanged without revealing the initial observation from nodes. Analysis of convergence time on the complete graph and line graph is given in the original paper in \cite{Kashyap}, and an $O(N^5)$ bound was derived in \cite{Zhu} by creating a random walk model.  

In this paper, we start with an analysis of convergence time of the distributed binary voting problem. We construct a biased lazy random walk model for this random process. We improve the upper bound on the expected convergence time in \cite{Shang} from $O(N^4\log N)$ to $O(N^3\log N)$. We then extend our results to the multi-level quantized consensus problem with the use of Lyapunov functions \cite{Zhu}\cite{Kashyap}. By utilizing the well-known relation between commuting time of a random walk and electric networks \cite{Aldous}, we derive an upper bound on the hitting time of a biased random walk. Several coupled Markov processes are then constructed to help the analysis. We improve the state of art bound in \cite{Zhu} from $O(N^5)$ to $O(N^3\log N)$. 

The contribution of this paper is as follows:
\begin{itemize}
\item A polynomial upper bound of $O(N^3\log N)$ for the quantized consensus algorithm. It is, to the best knowledge of the authors, the tightest bound in literature for the quantized consensus algorithm proposed in \cite{Benezit}\cite{Kashyap}. We use the degree of nodes on the shortest path  on the graph to improve the bound on the hitting time of the biased random walk.       
\item The analysis for arbitrary graphs is extended to a tighter bound for certain network topologies by computing the effective resistance between a pair of nodes on the graph. This is attractive because we can then apply results from algebraic graph theory \cite{Beineke}\cite{Godsil} to compute the effective resistance easily on the given graph structure.
\end{itemize}
The remainder of this paper is organized as follows. Section 2 describes the algorithm proposed in \cite{Benezit} and \cite{Kashyap}, and formulates the convergence speed problem. In Section 3, we derive our polynomial bound for this class of algorithms. In Section 4, we give examples on how to derive an upper bound on the given topology of the network, and simulation results are provided to justify the analysis. We provide our conclusions in Section 5.

%\begin{figure}[!t]
%\centering
%\centerline{\includegraphics[width=6.5cm]{sensors.jpg}}
%\caption{{\em Sensor nodes deployed in decision-making. \cite{Shang}}}
%\label{sensor}
%\end{figure}

\section{Problem Statement}
\label{sec:problem}

A network is represented by a connected graph $\mathcal{G = (V,E)}$, where $\mathcal{V} =\{1,2,...,N\}$ is the set of nodes and $\mathcal{E}$ is the set of edges. $(i,j)\in \mathcal{E}$ if nodes $i,j$ can communicate with each other. $\mathcal{N}_i$ is the set of neighbors of node $i$. 

Consider a network of $N$ nodes, labeled 1 through $N$. As proposed in \cite{Boyd}\cite{Kashyap}\cite{Benezit}, each node has a clock which ticks according to a rate 1 exponential distribution. By the superposition property for  the exponential distribution, this set up is equivalent to a single global clock with a rate $N$ exponential distribution ticking at times $\{Z_k\}_{k\ge0}$. The communication and update of  states only occur at  $\{Z_k\}_{k\ge0}$. When the clock of node $i$ ticks, $i$ randomly chooses a neighbor $j$ from the set $\mathcal{N}_i$.  We say edge $(i,j)$ is activated. 

In the rest of this section, we will describe the distributed binary voting consensus algorithm \cite{Benezit} and quantized consensus algorithm \cite{Kashyap}. We are interested in the performance of this class of algorithms on arbitrary graphs.

\subsection{Binary Voting Consensus}
\label{binary}

Initially, each node on the connected graph $\mathcal{G}$ has a vote, \emph{strong positive} or \emph{strong negative} (or no vote at all). Assuming that a majority opinion exists, the objective for the binary voting consensus problem is to have each node settle on the majority in a distributed manner.  

Let $S^{(i)}(t)$ denote the state of node $i$ at time $t$. $S^{(i)}(t) \in \{S^+, S^-, W^+, W^-\}$, representing \emph{strong positive}, \emph{strong negative}, \emph{weak positive}, and \emph{weak negative} respectively, where $S^\pm = \pm2$ and  $W^\pm = \pm1$.  For all $i \in \mathcal{V}$, $S^{(i)}(0)$ is initialized to the corresponding \emph{strong positive} or \emph{strong negative}. If $i$ does not have a vote at $t = 0$, it is randomly initialized to either \emph{weak} opinion. When 
 two nodes $i$ and $j$ with opposite \emph{strong opinions} exchange information, they both update to \emph{weak opinions}. Further update rules are as follows:

\begin{enumerate}
  \item If $S^{(i)}(t)=S^{(j)}(t), \\S^{(i)}(t+1)=S^{(j)}(t+1)=S^{(i)}(t);$
  \item If $|S^{(i)}(t)|>|S^{(j)}(t)|$ and  $S^{(i)}(t)\cdot S^{(j)}(t)<0, \\S^{(i)}(t+1)=-S^{(j)}(t), \;S^{(j)}(t+1)=S^{(i)}(t)$, and vice versa;
  \item If $|S^{(i)}(t)|>|S^{(j)}(t)|$ and  $S^{(i)}(t)\cdot S^{(j)}(t)>0, \\S^{(i)}(t+1)=S^{(j)}(t), \;S^{(j)}(t+1)=S^{(i)}(t)$, and vice versa;
  \item If $S^{(i)}(t)=-S^{(j)}(t), \\S^{(i)}(t+1)=\textrm{sign} \left(S^{(j)}(t)\right),\\S^{(j)}(t+1)=\textrm{sign}\left(S^{(i)}(t)\right).$
 \end{enumerate}
 
 \begin{figure}[t]
\centering
\centerline{\includegraphics[width=6.5cm]{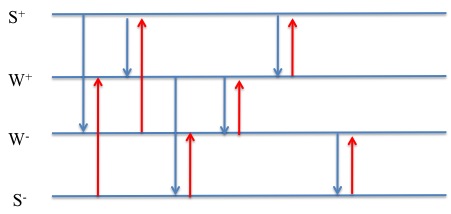}}
\caption{{\em Update rules for distributed binary vote} \cite{Benezit}. \em{The figure shows update principles: when opposite ``strong opinion"s meet, they both turn into ``weak opinion"s; ``strong opinion" affects ``weak opinion"; and swap principle.}}
\label{voting}
\end{figure}

The update rules are illustrated in Fig. \ref{voting}. Let $|S^+|$ denote the number of the \emph{strong positive} opinions and $|S^+(t)|$ denote the number of the \emph{strong positive} opinions at time $t$. Note that this algorithm supposes that there is an odd number of nodes in the network, in order to guarantee convergence regardless of initial votes of nodes. 

\begin{mydef}[Convergence on Binary Voting Consensus]
A binary voting reaches convergence if all states of nodes on the graph are positive or all states are negative.
\end{mydef}

A quick validation for this algorithm: we notice that the $S^+$ and $S^-$ will only annihilate each other when they meet, otherwise they just take random walks on the graph.  So only the majority \emph{strong opinions} will be left on the graph in the end. That is the reason why randomly assigning weak opinions to nodes with no initial vote does not affect the convergence to the majority opinion. We also notice that \emph{strong opinions} can influence \emph{weak opinions} as shown in Fig. \ref{voting}. Eventually all agents will take the sign of the majority  \emph{strong opinions}. Because the graph has finite size, and this Markov chain  has finite states, convergence will happen in finite time almost surely. 

\subsection{Quantized Consensus}
\label{sub:qc}
Without loss of generality, let us assume that all nodes hold integer values and the quantization is 1. Let $Q^{(i)}(t)$ denote the integer value of node $i$ at time $t$, with $Q^{(i)}(0)$ denoting the initial values. Define 
\begin{equation}
\label{qsum}
Q_{\rm{sum}} = \sum_{i  = 1} ^N Q^{(i)}(0).
\end{equation}
Let $Q_{\rm{sum}}$ be written as  $qN + r$, where $ 0 \le r < N$. Then the mean of the initial value in the network $\frac{1}{N} Q_{\rm{sum}} \in [q, q+1)$. Thus either $q$ or $q+1$ is an acceptable integer value for quantized average consensus (if the quantization level is 1). 

\begin{mydef}[Convergence on Quantized Consensus]
A quantized consensus reaches convergence at time $t$, if for any node $i$ on the graph,  $Q^{(i)}(t) \in \{q, q+1\}$.
\end{mydef}

There are a few properties that are desired for the quantized consensus algorithm: 
\begin{itemize}
\item \em{Sum conservation}:  
\begin{equation}
\sum_{i  = 1} ^N Q^{(i)}(t) = \sum_{i  = 1} ^N Q^{(i)}(t+1).
\end{equation}

\item {Variation non-increasing}:  \rm{if two nodes $i$, $j$ exchange information, }
\begin{equation}
|Q^{(i)}(t+1) - Q^{(j)}(t+1)| \le |Q^{(i)}(t) - Q^{(j)}(t)|.
\end{equation}
\end{itemize}

When two nodes $i$ and $j$ exchange information, without loss of generality, suppose that $Q^{(i)}(t)  \le Q^{(j)}(t)$. They follow the simple update rules below:
\begin{enumerate}
\item If $Q^{(j)}(t) - Q^{(i)}(t) \ge 2$, a \emph{non-trivial exchange} occurs:
$$
Q^{(i)}(t+1) = Q^{(i)}(t) + 1, Q^{(j)}(t+1) = Q^{(j)}(t) - 1.
$$
\item If $Q^{(j)}(t) - Q^{(i)}(t) \le 1$, a \emph{trivial exchange} occurs:

$$Q^{(i)}(t+1) = Q^{(j)}(t), Q^{(j)}(t+1) = Q^{(i)}(t).$$

\end{enumerate}

Similar to the argument in Section \ref{binary}, we can view this random process as a finite state Markov chain. Because the variation decreases whenever there is a non-trivial exchange,  convergence will be reached in finite time almost surely.  

\emph{Remark 1:} In this quantized consensus algorithm, we define convergence to be when all nodes reach two consecutive states. However, This definition cannot be used to solve the binary voting problem. If all nodes converge to $W^+$ and $W^-$ states, no conclusion can be made on the majority opinion.

\emph{Remark 2:} In this section, the update rules allow the node values to change by at most 1. This is relevant to load-balancing systems where only one value can be exchanged in the channel at a time due to the communication limit \cite{Kashyap}. Adjustments can be made for this class of quantized consensus algorithms, e.g. when two nodes exchange information, both nodes can update their value to the mean of the two. The analysis on the convergence time remains similar.

\section{Convergence Time Analysis}\label{main}
The main results of this work are the following theorems:
\begin{thm}
\label{mainth}
For a connected network of N nodes, an upper bound for the expected convergence time of the binary voting consensus algorithm is $\mathcal{O}(N^3\log(N))$.
\end{thm}
\begin{thm}
For a connected network of N nodes, an upper bound for the expected convergence time of the quantized consensus algorithm is $\mathcal{O}(N^3\log(N))$.
\end{thm}

We use the analogy of electric networks and random walks to derive the upper bound. Since the clock setting and edge selection strategies are the same in the binary voting algorithm and quantized consensus algorithm, their information exchange processes can be coupled. Hence the meeting time (defined below) is equal in both algorithms. Before deriving the bound on the convergence time, we first provide some definitions and notation that we will use and prove some useful lemmas in   Section \ref{def} and Section \ref{mt}. 

\subsection{Definition and Notation}\label{def}
\begin{mydef}[Hitting Time]
For a graph $\mathcal{G}$ and a specific random walk, let $\mathcal{H}{(i,j)}$ denote the expected number of steps a random walk beginning at $i$ must take before reaching $j$. Define the ``hitting time'' of $\mathcal{G}$ by $\mathcal{H(G)} = \max_{i,j}\mathcal{H}(i,j)$.
\end{mydef}

\begin{mydef}[Meeting Time]

Consider two random walkers placed on $\mathcal{G}$. At each tick of the clock, they move according to some joint probability distribution. Let $\mathcal{M}{(i,j)}$ denote the expected time for the two walkers to meet at the same node or to cross each other through the same edge (if they move at the same time). Define the ``meeting time" of $\mathcal{G}$ by $\mathcal{M(G)} = \max_{i,j}\mathcal{M}(i,j)$.
\end{mydef}

Define a \emph{simple random walk} on $\mathcal{G}$, $\mathcal{X}_S$, with transition matrix $P^{S} = (P_{ij})$ as follows:
\begin{itemize}
\item $P^S_{ii}: =0$ for $\forall i \in \mathcal{V}$,
\item $P^S_{ij}: =\frac{1}{|\mathcal{N}_i|}$ for $(i, j) \in \mathcal{E}$.  
\end{itemize}
$\mathcal{N}_i$ is the set of neighbors of node $i$ and $|\mathcal{N}_i|$ is the degree of node $i$.

Define a \emph{natural random walk $\mathcal{X}_N$} with transition matrix $P^{N} = (P_{ij})$ as follows:
\begin{itemize}
\item $P^N_{ii}=1-\frac{1}{N}$ for $\forall i \in \mathcal{V}$,
\item $P^N_{ij}=\frac{1}{N|\mathcal{N}_i|}$ for $(i, j) \in \mathcal{E}$.
\end{itemize}

Define a \emph{biased random walk $\mathcal{X}_B$} with transition matrix $P^B = (P_{ij})$ as follows:
\begin{itemize}
\item $P^B_{ii}: =1-\frac{1}{N} - \sum_{k\in \mathcal{N}_i}\frac{1}{N|\mathcal{N}_k|}$ for $\forall i \in \mathcal{V}$,
\item $P^B_{ij}: =\frac{1}{N}\left(\frac{1}{|\mathcal{N}_i|} + \frac{1}{|\mathcal{N}_j|}\right)$ for $(i, j) \in \mathcal{E}$.
\end{itemize}

\subsection{Hitting Time and Meeting Time on Weighted Graph}\label{mt}

In this class of algorithms, we label the initial observations by the nodes as $\alpha_1, \alpha_2, ..., \alpha_N$. Before any two observations $\alpha_m, \alpha_n$ meet each other, they take random walks on the graph $\mathcal{G}$. Their marginal transition matrices are both $P^B$. It may be tempting to think that they are taking the \emph{natural random walks} as stated in \cite{Zhu}. Upon closer inspection, we find that there are two sources stimulating the random walk from $i$ to $j$, for all $(i,j)\in \mathcal{E}$: one is active, from the node $i$, $P^1_{ij} = P^N_{ij}$; the other one is passive, from its neighbor $j$, $P^2_{ij} = P^N_{ji}$. Thus $P_{ij} = P^1_{ij} + P^2_{ij}$; i.e., the transitional matrix is actually $P^B$ instead of $P^N$. Denote this random process as $\mathcal{X}$. Because of the system settings, two random walks $\alpha_m, \alpha_n$ can only move at the same time if they are adjacent. Suppose $\alpha_m$ is at node $x$, and $\alpha_n$ is at node $y$. 

For $x \notin \mathcal{N}_y$,  and  $i \in \mathcal{N}_x$, we have
\begin{eqnarray}\nonumber 
&& P_{\mathcal{X}\textrm{joint}}(\alpha_m\textrm{   moves from } x \textrm{ to } i,\textrm{ } \alpha_n\textrm{ does not move}) \\ \nonumber &= &
P^B_{xi} - P_{\mathcal{X}\textrm{joint}}(\alpha_m\textrm{ moves from } x \textrm{ to } i\textrm{, } \alpha_n\textrm{ moves}) \\ 
& = &P^B_{xi}.
\label{bayes}
\end{eqnarray}
Similar for $P_{\mathcal{X}\textrm{joint}}(\alpha_n\textrm{   moves from } y \textrm{ to } j ,\textrm{ } \alpha_m\textrm{ does not move})$.
 Also, 
\begin{eqnarray} \nonumber
&& P_{\mathcal{X}\textrm{joint}}(\alpha_m\textrm{ does not move, } \alpha_n\textrm{ does not move}) \\
&=& 1 - \sum_{i\in \mathcal{N}_x}P^B_{xi} - \sum_{j\in \mathcal{N}_y}P^B_{yj}.
\end{eqnarray}
For $x \in \mathcal{N}_y$ and $i \neq y$ we have,  
%\begin{alignat}{3}
\begin{eqnarray}
\nonumber 
&& P_{\mathcal{X}\textrm{joint}}(\alpha_m\textrm{   moves from } x \textrm{ to } i,\textrm{ } \alpha_n\textrm{ does not move}) \\ \nonumber &= &
P^B_{xi} - P_{\mathcal{X}\textrm{joint}}(\alpha_m\textrm{ moves from } x \textrm{ to } i\textrm{, } \alpha_n\textrm{ moves}) \\ 
& = &P^B_{xi}.
\end{eqnarray}
\begin{eqnarray} 
 P_{\mathcal{X}\textrm{joint}}(\alpha_m\textrm{ moves to }y, \alpha_n \textrm{ moves to }x) = P^B_{xy}
 \label{meet}
\end{eqnarray}
\begin{eqnarray} 
\nonumber
&& P_{\mathcal{X}\textrm{joint}}(\alpha_m\textrm{ does not move, } \alpha_n\textrm{ does not move}) \\
&=& 1 - \sum_{i\in \mathcal{N}_x}P^B_{xi} - \sum_{j\in \mathcal{N}_y}P^B_{yj} + P^B_{xy}.
\label{stay}
\end{eqnarray}
%\end{alignat}

\begin{cor}
The biased random walk $\mathcal{X}_B$ is a reversible Markov process.
\end{cor}
\begin{proof}
Let $\pi$ be the stationary distribution of $\mathcal{X}_B$. It is easy to verify that
\begin{equation}
\pi_i = \frac{1}{N}
\end{equation}
 for all $i\in \mathcal{V}$.
Thus by the symmetry of $P^B$, $$\pi_iP^B_{ij} = \pi_jP^B_{ji}.$$
\end{proof}

\begin{lem} \label{lem:hitting}
In an arbitrary connected graph $\mathcal{G}$ with $N$ nodes, the hitting time of the biased random walk $\mathcal{X}_B$ satisfies $$\mathcal{H}_{P^B}(\mathcal{G}) < 3{N^3}.$$
\end{lem}
\begin{proof}
The biased random walk $\mathcal{X}_B$ defined above is a random walk on a weighted graph with weight
\begin{equation}\label{weight}
w_{ij}: = \frac{1}{N}\left(\frac{1}{|\mathcal{N}_i|} + \frac{1}{|\mathcal{N}_j|}\right) \textrm{ for } (i, j) \in \mathcal{E}.
\end{equation}
\begin{equation}
w_{ii} : = 1 - \sum_{j \in \mathcal{N}_i}w_{ij}.
\end{equation} 
\begin{equation}
w_i = \sum_{j\in \mathcal{V}}w_{ij} = 1, \; w = \sum_i{w_i} = N. 
\end{equation}
It is well-known that there is an analogy between a weighted graph and an electric network, where a wire linking $i$ and $j$ has conductance $w_{ij}$, i.e., resistance $r_{ij} = 1/w_{ij}$ \cite{Aldous}\cite{Nash}. And they have the following relationship
\begin{equation}\label{commute}
\mathcal{H}_{P^B}(x, y) + \mathcal{H}_{P^B}(y, x) = wr'_{xy},
\end{equation}
where $r'_{xy}$ is the effective resistance in the electric network between node $x$ and node $y$. Since the degree of any node is at most $N-1$, for $(i, j) \in \mathcal{E}$, 
\begin{eqnarray}\nonumber 
w_{ij} &=& \frac{1}{N}\left(\frac{1}{|\mathcal{N}_i|} + \frac{1}{|\mathcal{N}_j|}\right) \\ 
\nonumber & > & \frac{1}{N}\frac{1}{\min(|\mathcal{N}_i|, |\mathcal{N}_j|)} 
\end{eqnarray}

\begin{equation}
\label{h1}
r_{ij} < N\times \min(|\mathcal{N}_i|, |\mathcal{N}_j|)
\end{equation}

Consequently, $r'_{ij} \le r_{ij} <N\times \min(|\mathcal{N}_i|, |\mathcal{N}_j|)$. 

For all $x, y \in \mathcal{V}$, let $Q = ( q_1 = x, q_2, q_3, ..., q_{l-1}, q_l = y)$ be the shortest path on the graph connecting $x$ and $y$ . Now we claim that 
$$\sum^l _{k = 1}|\mathcal{N}_{q_k}| < 3N.$$

Since any node, say $u$ not lying on the shortest path can only be adjacent to at most three vertices on $Q$ (otherwise $u$ must be on the shortest path), we have
\begin{equation}
\label{h2}
\sum^l _{k = 1}|\mathcal{N}_{q_k}| \le 2l + 3(N - l) < 3N.
\end{equation}

By (\ref{h1}) and (\ref{h2}), we have 
\begin{equation}
r'_{xy} \le N \times \sum^l _{k = 1}|\mathcal{N}_{q_k}|  < 3N^2
\end{equation}

By (\ref{commute}), we have 
\begin{eqnarray}\nonumber 
\mathcal{H}_{P^B}(x,y) &<& \mathcal{H}_{P^B}(x, y) + \mathcal{H}_{P^B}(y, x) \\ \nonumber
&=& wr'_{xy}  \\ \nonumber
&<& N \times 3N^2  \\
&=& 3N^3.
\end{eqnarray}
This completes the proof.
\end{proof}
Note that this is an upper bound for arbitrary connected graphs. A tighter bound can be derived for certain network topologies. Examples will be given in Section \ref{sec: sim}.

\begin{cor}\label{circle}
$\mathcal{H}_{P^B}(x, y) + \mathcal{H}_{P^B}(y, z) + \mathcal{H}_{P^B}(z, x) = \mathcal{H}_{P^B}(x, z) + \mathcal{H}_{P^B}(z, y) + \mathcal{H}_{P^B}(y, x). $
\end{cor}
\begin{proof}
This is direct result from Lemma 2 in Chap 3 of Aldous-Fill's book \cite{Aldous} since $\mathcal{X}_B$ is reversible. 
\end{proof}

\begin{mydef}[Hidden Vertex]
A vertex $t$ in a graph is said to be hidden if for every other point in the graph,  $\mathcal{H}(t, v) \le\mathcal{H}(v, t)$. A hidden vertex is shown to exist for all reversible Markov chains in \cite{Coppersmith}.
\end{mydef} 

\begin{lem}\label{lem:meeting}
The meeting time of any two states on the network $\mathcal{G}$ following the random processes $\mathcal{X}$ in Section \ref{mt} is less than $4\mathcal{H}_{P^B}{(\mathcal{G})}$.
\end{lem}
\begin{proof}
In order to prove the lemma, we construct a coupling Markov chain, $\mathcal{X}'$ to assist the analysis. $\mathcal{X}'$ has the same joint distribution as $\mathcal{X}$  except (\ref{meet}) and (\ref{stay}).
\begin{eqnarray} 
 P_{\mathcal{X}'\textrm{joint}}(\alpha_m\textrm{, }\alpha_n \textrm{ meet at }x \textrm{ or }y) = 2P^B_{xy}
 \label{meet2}
\end{eqnarray}
\begin{eqnarray} \nonumber
&& P_{\mathcal{X}'\textrm{joint}}(\alpha_m\textrm{ does not move, } \alpha_n\textrm{ does not move}) \\
&=& 1 - \sum_{i\in \mathcal{N}_x}P^B_{xi} - \sum_{j\in \mathcal{N}_y}P^B_{yj}.
\label{stay2}
\end{eqnarray}
First, we show that the meeting time of two random walkers following $\mathcal{X}'$ is less than $2
\mathcal{H}_{P^B}{(\mathcal{G})}$. 

For convenience, we adopt the following notation: if $f(\cdot)$ is a real valued function on the vertex of the graph, then $f(\bar{v})$ is the weighted average of $f(u)$ over all neighbors $u$ of $v$, weighted according to the edge weights.

Similar as in \cite{Zhu}\cite{Coppersmith}, define a \emph{potential function}
\begin{equation}
\phi(x,y) := \mathcal{H}_{P^B}(x,y) + \mathcal{H}_{P^B}(y, t) - \mathcal{H}_{P^B}(t,y),
\end{equation}
 where $t$ is a hidden vertex on the graph. By Corollary \ref{circle}, $\phi(x,y)$ is symmetric, i.e. $\phi(x,y) = \phi(y,x)$. By the definition of meeting time, $\mathcal{M}$ is also symmetric, i.e. $\mathcal{M}(x,y) = \mathcal{M}(y,x)$. Next we use $\phi$ to bound the meeting time. 
 
 By the definition of hitting time, for $x \neq y$ we have
\begin{eqnarray} \nonumber
&&\mathcal{H}_{P^B}(x, y) \\ \nonumber
&=& 1 + P^B_{xx}\mathcal{H}_{P^B}(x,y) + \sum_{i\in \mathcal{N}_x}P^B_{xi}\mathcal{H}_{P^B}(i,y)\\
&=&1 + w_{xx}\mathcal{H}_{P^B}(x,y) + \sum_{i\in \mathcal{N}_x}w_{xi}\mathcal{H}_{P^B}(i,y), 
\end{eqnarray}
i.e.,
\begin{eqnarray}\nonumber
\mathcal{H}_{P^B}(x, y) &=& \frac{1}{\sum_{i\in \mathcal{N}_x}w_{xi}} + \frac{\sum_{i\in \mathcal{N}_x}w_{xi}\mathcal{H}_{P^B}(i,y)}{\sum_{i\in \mathcal{N}_x}w_{xi}} \\
&=& \frac{1}{\sum_{i\in \mathcal{N}_x}w_{xi}} + \mathcal{H}(\bar{x},y).
\end{eqnarray}
So for $x \neq y$,
\begin{equation} \label{phi}
\phi(x,y) = \frac{1}{\sum_{i\in \mathcal{N}_x}w_{xi}} + \phi(\bar{x},y).
\end{equation}
\begin{eqnarray}\nonumber 
\mathcal{M}_{\mathcal{X}'}(x, y)  &= & 1 + \left( 1 - \sum_{i \in \mathcal{N}_x}P^B_{xi} - \sum_{j \in \mathcal{N}_y}P^B_{yj}\right)\mathcal{M}_{\mathcal{X}'}(x,y) \\ \nonumber 
& + & \sum_{i \in \mathcal{N}_x}P^B_{xi}\mathcal{M}_{\mathcal{X}'}(i,y) \\
& + & \sum_{j \in \mathcal{N}_y}P^B_{yj}\mathcal{M}_{\mathcal{X}'}(x,j).
\label{universal}
\end{eqnarray}
Note that   (\ref{universal}) also holds for $x\in \mathcal{N}_y$. We now have
\begin{eqnarray}\nonumber 
&& \left( \sum_{i \in \mathcal{N}_x}P^B_{xi} + \sum_{j \in \mathcal{N}_y}P^B_{yj}\right)\mathcal{M}_{\mathcal{X}'}(x,y)  
\\ & = &  1 +  \sum_{i \in \mathcal{N}_x}P^B_{xi}\mathcal{M}_{\mathcal{X}'}(i,y) 
 +  \sum_{j \in \mathcal{N}_y}P^B_{yj}\mathcal{M}_{\mathcal{X}'}(x,j).
 \label{ineq}
\end{eqnarray}
  (\ref{ineq}) shows that at least one of the two inequalities below holds:
\begin{equation}
\mathcal{M}_{\mathcal{X}'}(x, y) > \frac{\sum_{i \in \mathcal{N}_x}P^B_{xi}\mathcal{M}_{\mathcal{X}'}(i,y)}{\sum_{i \in \mathcal{N}_x}P^B_{xi}} = \mathcal{M}_{\mathcal{X}'}(\bar{x},y)
\end{equation}
\begin{equation}
\mathcal{M}_{\mathcal{X}'}(x, y) > \frac{\sum_{j \in \mathcal{N}_y}P^B_{yj}\mathcal{M}_{\mathcal{X}'}(x,j)}{\sum_{j \in \mathcal{N}_y}P^B_{yj}} = \mathcal{M}_{\mathcal{X}'}(x,\bar{y})
\label{yineq}
\end{equation}
Without loss of generality, suppose that (\ref{yineq}) holds (otherwise, we can prove the other way around). From (\ref{ineq}), we have
\begin{eqnarray}\nonumber 
&& \sum_{i \in \mathcal{N}_x}P^B_{xi} \mathcal{M}_{\mathcal{X}'}(x,y)  
  =    1 +  \sum_{i \in \mathcal{N}_x}P^B_{xi}\mathcal{M}_{\mathcal{X}'}(i,y) \\
 & + &  \sum_{j \in \mathcal{N}_y}P^B_{yj}\mathcal{M}_{\mathcal{X}'}(x,j) - \sum_{j \in \mathcal{N}_y}P^B_{yj}\mathcal{M}_{\mathcal{X}'}(x,y).
\end{eqnarray}
i.e.,
\begin{eqnarray}\nonumber  
\mathcal{M}_{\mathcal{X}'}(x,y)  
   & = &   \frac{1}{\displaystyle\sum_{i \in \mathcal{N}_x}P^B_{xi}} +  \mathcal{M}_{\mathcal{X}'}(\bar{x},y) \\\nonumber
& + &  \frac{\displaystyle\sum_{j \in \mathcal{N}_y}P^B_{yj}\left(\mathcal{M}_{\mathcal{X}'}(x,\bar{y}) - \mathcal{M}_{\mathcal{X}'}(x,y)\right)}{\displaystyle\sum_{i \in \mathcal{N}_x}P^B_{xi}} \\
& <  & \frac{1}{\displaystyle\sum_{i \in \mathcal{N}_x}w_{xi}} + \mathcal{M}_{\mathcal{X}'}(\bar{x},y).
\label{lastineq}
\end{eqnarray}
Now we claim that $\mathcal{M}_{\mathcal{X}'}(x,y) \le \phi(x,y)$. Suppose it is not the case. Let $\beta = \max_{x,y}\{\mathcal{M}_{\mathcal{X}'}(x,y) - \phi(x,y)\}$. Among all the pairs $x, y$ realizing $\beta$, choose any pair.
It is clear that $x \neq y$, since $\mathcal{M}_{\mathcal{X}'}(x,x) = 0 \le \phi(x,x)$. By (\ref{phi}) and (\ref{lastineq}), 
\begin{eqnarray}\nonumber
\mathcal{M}_{\mathcal{X}'}(x,y) &=& \phi(x,y) + \beta \\ \nonumber
&=& \frac{1}{\sum_{i\in \mathcal{N}_x}w_{xi}} + \phi(\bar{x}, y) + \beta\\  \nonumber
&\ge& \frac{1}{\sum_{i\in \mathcal{N}_x}w_{xi}} + \mathcal{M}_{\mathcal{X}'}(\bar{x},y) \\
& > &\mathcal{M}_{\mathcal{X}'}(x,y).
\end{eqnarray}
This is a contradiction. Thus $\mathcal{M}_{\mathcal{X}'}(\mathcal{G}) <  \phi(x,y) < 2\mathcal{H}_{P^B}{(\mathcal{G})}$.

Now we are ready to complete the proof of Lemma \ref{lem:meeting}. We couple the Markov chains $\mathcal{X}$ and $\mathcal{X'}$ so that they are equal until the two random walkers become neighbors. Note that half of the time when the walkers in $\mathcal{X}'$ meet, they do not meet in  $\mathcal{X}$ , but stay in the same position. We claim that $\mathcal{M}_{\mathcal{X}}(\mathcal{G}) \leq 2\mathcal{M}_{\mathcal{X}'}(\mathcal{G})$. 

In the random process $\mathcal{X}'$, when two random walkers $m$, $n$ meet, instead of finishing the process, we let them cross and continue the random walks according to $P_{\mathcal{X}'\textrm{joint}}$. The expected length of each cross is less than or equal to $\mathcal{M}_{\mathcal{X}'(\mathcal{G})}$. At each cross, the random process $\mathcal{X}$ finishes with a probability of 1/2, independently. Thus for any $x, y \in \mathcal{V}$ we have 
\begin{equation}
\mathcal{M}_{\mathcal{X}}(x,y) \leq \sum_{i = 1}^{\infty}\left(\frac{1}{2}\right)^ii\mathcal{M}_{\mathcal{X}'}(\mathcal{G}) = 2\mathcal{M}_{\mathcal{X}'}(\mathcal{G}).
\end{equation}

This completes the proof.
\end{proof}

Now we are ready to prove Theorems 1 and 2 in the next two subsections.

\subsection{An Upper Bound on Binary Consensus}\label{ana}

Without loss of generality, let us suppose that in the initial setting more nodes hold \emph{strong positive} opinions ($S^+$). As briefly analyzed in Section \ref{binary}, the process undergoes two stages: the depletion of $S^-$ and the depletion of $W^-$. By our assumption, 
\begin{equation}
\label{ }
|S^+(0)| > |S^-(0)|
\end{equation}
and 
\begin{equation}
\label{ }
|S^+(0)| + |S^-(0)| = N,
\end{equation} 
where $N$ is the number of nodes on the graph.  

According to the update rules in Section \ref{binary}, we have
\begin{equation}
\label{ }
|S^+(t)| - |S^-(t)| = |S^+(0)|-|S^-(0)|.
\end{equation}

Let $T_1$ and $T_2$ denote the maximum expected time it takes for Stage 1 and Stage 2 to finish. In the first stage, two opposite \emph{strong opinions} annihilate when an edge between them is activated. Otherwise they take biased random walks on the graph $\mathcal{G}$.  
In the second stage, the remaining $|S^+(0)|-|S^-(0)|$ \emph{strong positive}s take random walks over graph $\mathcal{G}$, transforming \emph{weak negative} into \emph{weak positive}. 

Let $CT_{\mathcal{G}}(v)$ denote the expected time for a random walker starting from node $v$ to meet all other random walkers who are also taking random walks on the same graph but starting from different nodes. Define 
$$CT(\mathcal{G}) = \max_{v \in \mathcal{V}} CT_{\mathcal{G}}(v).$$

\begin{cor}\label{max}
Let $\mathcal{M}_{\mathcal{X}}(\mathcal{G})$ be the meeting time of the biased random walk $\mathcal{X}$ defined in Section \ref{binary}. Then
\begin{equation}
\label{ }
CT(\mathcal{G}) = \mathcal{O}(\mathcal{M}_{\mathcal{X}}(\mathcal{G})\log N).
\end{equation}\end{cor}
\begin{proof}
Since there are no more than $N$ consecutive meetings, we can easily get a union bound for $CT(\mathcal{G})$, which is $N\mathcal{M}_{\mathcal{X}}(\mathcal{G})$.

In order to obtain a tighter bound for $CT(\mathcal{G})$, we divide the random walk into $\ln N$ periods of length $k\mathcal{M}_{\mathcal{X}}(\mathcal{G})$ each, where $k$ is a constant. Let $a$ be the ``special" random walker trying to meet all other random walkers. For any period $i$ and any other random walker $v$, by the Markov inequality, we have
\begin{eqnarray}\nonumber
&&\Pr(\textrm{$a$ does not meet $v$ during period $i$})  \\ \nonumber
&\le& \frac{\mathcal{M}_{\mathcal{X}}(\mathcal{G})}{k\mathcal{M}_{\mathcal{X}}(\mathcal{G})} \\
&=& \frac{1}{k}
\end{eqnarray}
so
\begin{eqnarray}\nonumber
&&\Pr(\textrm{$a$ does not meet $v$ during any period})  \\
&\le& \left( \frac{1}{k}\right)^{\ln N} = N^{-\ln k}
\end{eqnarray}
If we take the union bound, 
\begin{eqnarray}\nonumber
&&\Pr(\textrm{$a$ doesn't meet some walker during any period}) \\
 &\le&  N\cdot N^{-\ln k}.
\end{eqnarray}

Conditioning on whether or not the walker $a$ has met all other walkers after all $k\mathcal{M}_{\mathcal{X}}(\mathcal{G})\ln N$ steps, and using the previous $N\mathcal{M}_{\mathcal{X}}(\mathcal{G})$ upper bound, we have

\begin{eqnarray}\nonumber
CT(\mathcal{G}) &\le& k\mathcal{M}_{\mathcal{X}}(\mathcal{G})\ln N + N\cdot N^{-\ln k}\cdot N\mathcal{M}_{\mathcal{X}}(\mathcal{G})\\ 
&=& k\mathcal{M}_{\mathcal{X}}(\mathcal{G})\ln N +  N^{2-\ln k}\mathcal{M}_{\mathcal{X}}(\mathcal{G})
\end{eqnarray}

When $k$ is sufficiently large, say $k \ge e^6$, the second term is small, so
\begin{equation}
\label{ }
CT(\mathcal{G}) < (k + 1)\mathcal{M}_{\mathcal{X}}(\mathcal{G})\ln N.
\end{equation}
This completes the proof. 
\end{proof}

\begin{proof}[Proof of Theorem 1]
In order to analyze Stage 1, we can construct a coupled Markov process. When two different strong opinions meet, instead of following the rules to change into weak opinions, they just keep their states and keep moving along the same path they would have as weak opinions. This process is over when every strong opinion has met all other opposite strong opinions, by when Stage 1 must have finished, i.e. before at most $N^2/4$ such meetings. The rest of the proof just follows from Corollary \ref{max}, except we divide the random walks into $\ln(N^2/4)$ periods of length $k\mathcal{M}_{\mathcal{X}}(\mathcal{G})$ instead of $\ln N$. Hence we have $T_1  \le 2CT(\mathcal{G})$. $T_2  \le CT(\mathcal{G})$ follows from the fact that there are at most $N-1$ meetings for a single strong opinion to meet all the weak opinions to ensure convergence. By Lemma \ref{lem:hitting} , Lemma \ref{lem:meeting} and Corollary \ref{max} , both $T_1$ and $T_2$ are $O(N^3\log N)$.

Hence the upper bound of convergence time of binary voting is thus $O(N^3\log N)$.
\end{proof}

\subsection{An Upper Bound on Quantized Consensus} \label{boundqc}
Recall that a non-trivial exchange in quantized consensus happens when the difference in values at the nodes is greater than 1. 

Let $\mathbf{Q}(t)$ denote a vector of values all nodes holding at time $t$. Set $\bar{Q} = Q_{\rm{sum}}/N$, where $Q_{\rm{sum}}$ is defined in   (\ref{qsum}).

We construct a Lyapunov function $L_{\bar{Q}}$ \cite{Zhu}\cite{Kashyap}\cite{Nedic} as:

\begin{equation}
L_{\bar{Q}}(\mathbf{Q}(t)) = \sum_{i = 1}^N \left(Q_i(t) - \bar{Q}\right)^2.
\end{equation} 

Let $m = \min_i Q_i(0)$ and $M = \max_i Q_i(0)$. It is easy to see that $L_{\bar{Q}}(\mathbf{Q}(0)) \le \frac{(M-m)^2N}{4}$. Equality holds when half of the values are $M$ and others are $m$.

\begin{cor}
\label{cor:lfunc}
In a non-trivial exchange, $$L_{\bar{Q}}(\mathbf{Q}(t)) \ge L_{\bar{Q}}(\mathbf{Q}(t+1)) + 2.$$
\end{cor}

\begin{proof}
A non-trivial exchange follows the first update rule of quantized consensus algorithm in Section \ref{sub:qc}. 

Suppose $Q_i(t) = x_1$ and $Q_j(t) = x_2$ have a non-trivial exchange at time $t$, and the rest of the values stay unchanged. Without loss of generality, let $x_1 \le x_2 - 2$.  We have

\begin{eqnarray} \nonumber
&&L_{\bar{Q}}(\mathbf{Q}(t)) - L_{\bar{Q}}(\mathbf{Q}(t+1)) \\ \nonumber
&=& x_1^2 +   x_2^2 - (x_1 + 1)^2 - (x_2 - 1)^2 \\ 
&=& 2(x_2 - x_1) - 2 \ge 2.
\end{eqnarray}
\end{proof}

\begin{proof}[Proof of Theorem 2]
Corollary \ref{cor:lfunc} shows that the Lyapunov function is decreasing. The convergence of quantized consensus must be reached after at most $\gamma = \frac{(M-m)^2N}{8}$ non-trivial exchanges. Similar to the analysis for the binary voting algorithm, when every random walker has met each of the other random walkers  $\gamma$ times, all the non-trivial exchanges must have finished.  By Corollary \ref{max}, this process finishes in $O(N^3\log N)$ time. 
\end{proof}

\section{Simulation Results}
\label{sec: sim}
In this section, we give examples of star networks, line graphs and lollipop graphs in order to show how to use the analysis in Section \ref{main} for the particular graphs with known topologies. Simulation results are provided to validate the analysis. 
We also simulate the distributed process on Erd\"os-R\'enyi random graph in order to get some insight on how the algorithm performs on a random graph.

%more graph topologies, such as line path graph, Erd\"os-R\'enyi random graph, and lollipop graph to justify the $O(N^4\log N)$ upper bound. 

\subsection{Star Networks}
\begin{figure}[!t]
\centering
\centerline{\includegraphics[width=3.5cm]{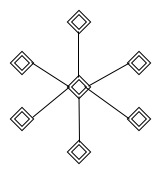}}
\caption{{\em A star network.}}
\label{star}
\end{figure}

%\subsection{Star Networks}
Star networks are a common network topology. A star network $\mathcal{S}$ of $N$ nodes has one central hub and $N-1$ leaf nodes, as shown in Fig. \ref{star}. Now let us derive an upper bound following the similar analysis in Section \ref{main}.

\subsubsection{Analysis}\label{starbound}
By   (\ref{weight}) in Section \ref{mt}, suppose that there is a star network of $N$ nodes, with the central hub denoted as $c$. For $\forall i,j \ne c$, we have
\begin{equation}
w_{ic} = w_{jc} = \frac{1}{N}\left(1 + \frac{1}{N-1}\right) = \frac{1}{N-1}.
\end{equation}
The equivalent resistance between any two leaf nodes $i$ and $j$ is
\begin{equation}
r'_{ij} = \frac{1}{w_{ic}} + \frac{1}{w_{jc}} = 2N - 2.
\end{equation}   
By the symmetry of the star network, it is easy to see that
\begin{equation}
\mathcal{H}_{P^B}(i, j) = \mathcal{H}_{P^B}(j, i).
\end{equation}
By   (\ref{commute}),
\begin{equation}
\mathcal{H}_{P^B}(\mathcal{S}) = \mathcal{H}_{P^B}(i,j) = N(N-1).
\end{equation}
Then following similar analysis in Section \ref{ana} and Section \ref{boundqc}, we can bound the convergence time of both binary and multi-level quantized consensus algorithms of a star network by $O(N^2\log N)$. 
\subsubsection{Simulations} \label{sim1}
We simulate the star networks with the number of nodes $N$ ranging from 21 to 481 , with intervals of 20, for both binary consensus and quantized consensus algorithms. For binary consensus, Initially, there are $\lceil N/2\rceil$ \emph{strong positive} and $\lfloor N/2\rfloor$ \emph{strong negative} nodes, i.e., $|S^+| - |S^-| = 1$. Those nodes communicate with each other following the protocol in Section \ref{binary}. The process finishes when consensus is reached. Simulation results on binary consensus are shown in Fig. \ref{fig:1}, and are indeed of order $O(N^2\log N)$, as analyzed above. 

For quantized consensus, we show two different initial settings: (1) $Q^{(i)}(0) = 2$, $Q^{(j)}(0) = 0$, $Q^{(k)}(0) = 1$, for $k \ne i, j$ and $i, j \ne c$, Fig. \ref{fig:2}; (2) Initial values of nodes are drawn uniformly from 1 to 100, Fig. \ref{fig:3}. Nodes on the graph exchange information according to the update rules in Section \ref{sub:qc}. 

We notice that quantized consensus algorithm converges faster than the binary consensus in the above setting, because a non-trivial exchange takes place whenever the difference of the values between the  selected nodes is greater than one. The Lyapunov value is non-increasing. However, in binary consensus, a strong negative opinion can influence the weak opinions before its annihilation. In the first setting in quantized consensus simulation, there is only one non-trivial exchange before reaching convergence, hence the convergence time is actually the meeting time of the graph. In the second setting, due to the uniform distribution of node values, the non-trivial exchange is more often than the first setting, because the special structure of a star network, central hub can balance the values of leaf nodes quickly.

Convergence time in all cases is the average of 20 rounds of simulations. 

%\begin{figure}[t]
%\centering
%\centerline{\includegraphics[width=.45\textwidth]{star.jpg }}
%\caption{{\em Average convergence time (green squares) versus the size of the star network. The blue solid line indicate $0.63N^2\log N$. }}
%\label{simulation}
%\end{figure}
%\begin{figure}[!t]
%\centering
%\centerline{\includegraphics[width=.5\textwidth]{simulation1.jpg}}
%\caption{{\em A star network.}}
%\label{simulation2}
%\end{figure}

%We plot the average convergence time (green squares) versus the size of star networks in Fig. \ref{simulation}. As in the figure indicated by the solid blue line, $O(N^2\log N)$ is indeed an upper bound. Fig. \ref{simulation} also indicates that this bound  is tight for a star network.

%\subsection{Complete Graphs}
%A complete graph $\mathcal{K}$ is a graph that every pair of distinct vertices is connected by a unique edge, as shown in Fig. \ref{complete}.

%
%\begin{figure}[!t]
%\centering
%\centerline{\includegraphics[width=4cm]{complete.jpg}}
%\caption{{\em A complete graph.}}
%\label{complete}
%\end{figure}

%\subsubsection{Analysis}
%
%\subsubsection{Simulations}
%
%\begin{figure*}[!t] % use float package if you want it here
%\centering
%\subfloat[A star network. ]{\label{star}\includegraphics[width=0.25\textwidth]{star.jpg}} \hfill
%\subfloat[A line graph. ]{\label{line}\includegraphics[width=0.3\textwidth]{line.jpg}} \hfill
%\subfloat[A lollipop graph. ]{\label{lollipop}\includegraphics[width=0.4\textwidth]{lollipop.pdf}}  \\
%\end{figure*}

\subsection{Line Graph}\label{sub:line}
\begin{figure}[!t]
\centering
\centerline{\includegraphics[width=.4\textwidth]{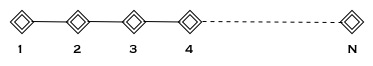}}
\caption{{\em A line graph.}}
\label{line}
\end{figure}

A line graph is a simple graph structure. Nodes in a line graph $\mathcal{L}$ are connected to one another in a line as shown in Fig. \ref{line}. Although a line graph is less realistic in applications, it serves as an example here to show that the convergence time can indeed reach $O(N^3\log N)$, thus to show the tightness of the derived bound. 

\subsubsection{Analysis}
In a line graph of $N$ nodes, for two adjacent nodes $i, j$ (not end points), we have
\begin{equation}
w_{ij} = \frac{1}{N}\left( \frac{1}{2} + \frac{1}{2}\right) = \frac{1}{N}.
\end{equation}
The equivalent resistance is 
\begin{equation}
r_{ij} = N.
\end{equation}
For two end points, similarly, both have resistance of $2/3 N$ with their neighbors. Thus the effective resistance between two end points $m, n$ of the line graph is $N^2 - \frac{5}{3}N$.  
By the symmetry of a line graph, the hitting time of the graph is 
\begin{equation}
\mathcal{H}_{P^B}(\mathcal{L}) = \mathcal{H}_{P^B}(m,n) = \frac{1}{2} N\left(N^2 - \frac{5}{3}N\right).
\end{equation}
The rest follows the analysis  in Section \ref{starbound}, an upper bound on the convergence time for binary and multi-level quantized consensus is $O(N^3\log N)$.

\subsubsection{Simulations}
Experiment settings for line graph are same as in Section \ref{sim1}. The results are plotted in Fig. \ref{fig:4} - Fig. \ref{fig:6}.

\subsection{Lollipop Graph}\label{sub:lollipop}

\begin{figure}[!t]
\centering
\centerline{\includegraphics[width=.4\textwidth]{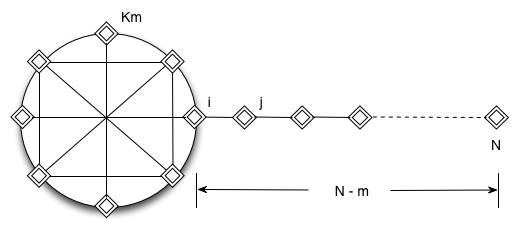}}
\caption{{\em A lollipop graph.}}
\label{lollipop}
\end{figure}

A lollipop graph is a line graph joined to a clique. Fig. \ref{lollipop} shows a lollipop graph $\mathcal{P}$ with $N$ nodes, $m$ of which form a clique $K_m$, and the rest of $N - m$ nodes connected to $K_m$ by node $i$ as a line. It is well known that a lollipop graph, when $m = \lfloor (2N+1)/3 \rfloor $, is the extremal graph for the maximum hitting time $O(N^3)$ of a \emph{simple random walk} \cite{Brightwell}\cite{Lovasz}. In a \emph{simple random walk} starting from $i$, the walker is very unlikely to go to $j$, compared with from $j$ to $i$. This  results in a latency factor of $N$ for a \emph{simple random walk} starting from the clique and going to the end of the line on the lollipop. For a \emph{natural random walk} $\mathcal{X_N}$ on $\mathcal{P}$, the hitting time is $O(N^4)$ \cite{Zhu}, because of the laziness of $\mathcal{X_N}$. However, it is not the case in a \emph{biased random walk}. Since $P^B_{ij} = P^B_{ji}$, it is equal likely that a random walker moves from $i$ to $j$ and from $j$ to $i$.  

For any two nodes $s, t$ on the clique $K_m$, and $m$ is $O(N),$
\begin{equation}
w_{st} = \frac{1}{N}\left(\frac{1}{m-1} + \frac{1}{m-1}\right) = \frac{2}{N(m-1)}.
\end{equation}
The effective resistance $r'_{st}$ of a clique is clearly less than $r_{st} = 1/w_{st} = O(N^2),$ and therefore the hitting time from any node on $K_m$ to $i$ is $O(N^3)$. Furthermore, the line on $\mathcal{P}$ has hitting time of $O(N^3)$, as analyzed in Section \ref{sub:line}. We then have
\begin{equation} 
\mathcal{H}_{P^B}(\mathcal{P}) = O(N^3).
\end{equation}
Similarly, an upper bound of convergence time is $O(N^3\log N)$ for lollipop graph. 

Experiment settings for a lollipop graph are the same as in Section \ref{sim1}. The results are plotted in Fig. \ref{fig:7} - Fig. \ref{fig:9}.

\subsection{Erd\"os-R\'enyi random graph}
%A \emph{line path graph} is a graph that nodes are connected in a line. 
In an Erd\"os-R\'enyi random graph $\mathcal{R}$, an edge is set between each pair of nodes independently with equal probability $p$. As one of the properties of Erd\"os-R\'enyi random graphs, when $p > \frac{(1+\epsilon)\log N}{N}$, the graph $\mathcal{R}$ will almost surely be connected \cite{Durrett}. $$E(\textrm{number of edges}) = 0.5N(N-1)p.$$ The diameter of Erd\"os-R\'enyi random graphs is rather sensitive to small changes in the graph, but the typical distance between two random nodes on the graph is
$d = \frac{\log N}{\log (pN)}$ \cite{Durrett}.

We created Erd\"os-R\'enyi random graphs by setting $p = 5\log N /N$, where $N$ ranged from $21$ to $481$, with an interval of $20$. Other settings are the same as in Section \ref{sim1}. Experiment results of Erd\"os-R\'enyi random graphs are shown in Fig. \ref{fig:10} - Fig \ref{fig:12}. It appears that the expected convergence time of binary consensus is on the order of $N^2\log N$, which is lower than the general upper bound of Theorem 1.
%
%Following is some informal analysis on the convergence speed. Since we know that the diameter of Erd\"os-R\'enyi random graphs is rather sensitive to small changes in the graph, we can take the typical distance \cite{Durrett} of two random nodes on a Erd\"os-R\'enyi random graph instead: 
%\begin{equation}
%d = \frac{\log N}{\log (pN)} = \frac{\log N}{\log (5\log N)}, 
%\end{equation}
%and 
%\begin{equation}
%\label{ }
%E(\textrm{degree of a node}) = p(N-1)\approx pN = 5\log N.
%\end{equation}
%For $(i,j) \in \mathcal{E}$, $w_{ij}\sim \frac{1}{N}\frac{2}{pN} = \frac{2}{5N\log N}$. Thus a typical effective resistance $$r' < \frac{1}{w_{ij}}d\sim$$

%\begin{figure}[!t]
%\centering
%\centerline{\includegraphics[width=.45\textwidth]{ER.jpg }}
%\caption{{\em Average convergence time versus the size of the Erd\"os-R\'enyi random graph. The blue solid line indicates $2N^2\log N$, and the red dash line indicates $2.3N^2\log N$.}}
%\label{simulation2}
%\end{figure}

\section{conclusions}
\label{sec: discussion}
In this paper, we use the theory of electric networks, random walks, and couplings of Markov chains to derive a polynomial bound on convergence time with respect to the size of the network, for a class of distributed quantized consensus algorithms \cite{Benezit}\cite{Kashyap}. We improve the state of art bound of $O(N^4\log N)$ for binary consensus and $O(N^5)$ for quantized consensus algorithms to $O(N^3\log N)$. Our analysis can be extended to a tighter bound for certain network topologies using the effective resistance analogy.  Our results provide insights to the performance of the binary and multi-level quantized consensus algorithms.

%\section{acknowledgement}
%This research was supported in part by the Center for Science of
%Information (CSoI), an National Science Foundation (NSF) Science and Technology Center, under grant
%agreement CCF-0939370, by NSF under the grant CCF-1116013,
%by the U.S. Army Research Office under grant
%number W911NF-07-1-0185, and by a research grant from Deutsche Telekom
%AG.

%\section{REFERENCES}
%\label{sec:ref}

\bibliographystyle{IEEEbib}
\bibliography{myrefs}

\begin{thebibliography}{10}

\bibitem{Boyd}
S.~Boyd, A.~Ghosh, B.~Prabhakar, and D.~Shah,
\newblock ``Randomized gossip algorithms,''
\newblock {\em IEEE Transactions on Information Theory}, pp. 2508--2530, 2006.

\bibitem{Zhu}
M.~Zhu and S.~Mart{\'\i}nez,
\newblock ``On the convergence time of asynchronous distributed quantized
  averaging algorithms,''
\newblock {\em IEEE Transactions on Automatic Control}, vol. 56, pp. 386--390,
  2011.

\bibitem{Benezit}
F.~B\'en\'ezit, P.~Thiran, and M.~Vetterli,
\newblock ``Intervalconsensus: From quantized gossip to voting,''
\newblock {\em Proc. of IEEE ICASP}, pp. 3661--3664, 2009.

\bibitem{Kashyap}
A.~Kashyap, T.~Basar, and R.~Srikant,
\newblock ``Quantized consensus,''
\newblock {\em Automatica}, pp. 1192--1203, 2007.

\bibitem{Cai}
K.~Cai and H.~Ishii,
\newblock ``Convergence time analysis of quantized gossip algorithms on
  digraphs,''
\newblock {\em in Proceedings 49th IEEE Conference on Decision and Control},
  p.~6, 2008.

\bibitem{Drief}
M.~Draief and M.~Vojnovic,
\newblock ``Convergence speed of binary interval consensus,''
\newblock {\em Proc. of IEEE INFOCOM}, pp. 1--9, 2010.

\bibitem{Du}
X.~Du, M.~Zhang, K.~E. Nygard, S.~Guizani, and H.~Chen,
\newblock ``Self healing sensor networks with distributed decision making,''
\newblock {\em International Journal of Sensor Networks}, vol. 2, no. 5/6, pp.
  289--298, July 2007.

\bibitem{Mossel}
E.~Mossel and G.~Schoenebeck,
\newblock ``Reaching consensus on social networks,''
\newblock {\em Innovations in Computer Science, ICS}, pp. 214--229, 2010.

\bibitem{Shang}
S.~Shang, P.~Cuff, S.~Kulkarni, and P.~Hui,
\newblock ``An upper bound on the convergence time for distributed binary
  consensus,''
\newblock {\em 15th International Conference on Information Fusion}, July 2012.

\bibitem{Nedic}
A.~Nedic, A.~Olshevsky, A.~Ozdaglar, and J.N. Tsitsiklis,
\newblock ``On distributed averaging algorithms and quantization effects,''
\newblock {\em IEEE Transactions on Automatic Control}, vol. 54, 2009.

\bibitem{Aldous}
D.Aldous and J.Fill,
\newblock {\em Reversible Markov Chains and Random Walks on Graphs},
\newblock http://www.stat.berkeley.edu/~aldous/RWG/book.html.

\bibitem{Beineke}
Lowell~W. Beineke and Robin~J. Wilson,
\newblock {\em Topics In Algebraic Graph Theory},
\newblock Cambridge University Press, 2004.

\bibitem{Godsil}
Chris Godsil and Gordon Royle,
\newblock {\em Algebraic Graph Theory},
\newblock Springer, 2001.

\bibitem{Nash}
C~Nash-Williams,
\newblock ``Random walk and electric currents in networks,''
\newblock {\em Proceedings of the Cambridge Philosophical Society}, vol. 55,
  no. 2, pp. 181--194, 1959.

\bibitem{Coppersmith}
D.~Coppersmith, P.~Tetali, and P.~Winkler,
\newblock ``Collisions among random walks on a graph,''
\newblock {\em SIAM J. on Discrete Mathematics}, vol. 6, pp. 363--374, 1993.

\bibitem{Brightwell}
G.~Brightwell and P.~Winkler,
\newblock ``Maximum hitting time for random walks on graphs,''
\newblock {\em Random Structures and Algorithms}, vol. 1, pp. 263--276, October
  1990.

\bibitem{Lovasz}
L.~Lov{\'a}sz,
\newblock ``Random walks on graphs: a survey,''
\newblock {\em Combinatorics, Paul Erd{\"o}s is Eighty}, vol. 2, pp. 1--46,
  1993.

\bibitem{Durrett}
Rick Durrett,
\newblock {\em Random Graph Dynamics},
\newblock Cambridge University Press, 2006.

\end{thebibliography}

\begin{figure*}[!t] % use float package if you want it here
\centering
\subfloat[Simulation results of average convergence time on binary consensus algorithm of star networks. The solid line indicates $0.63N^2\log N$. ]{\label{fig:1}\includegraphics[width=0.32\textwidth]{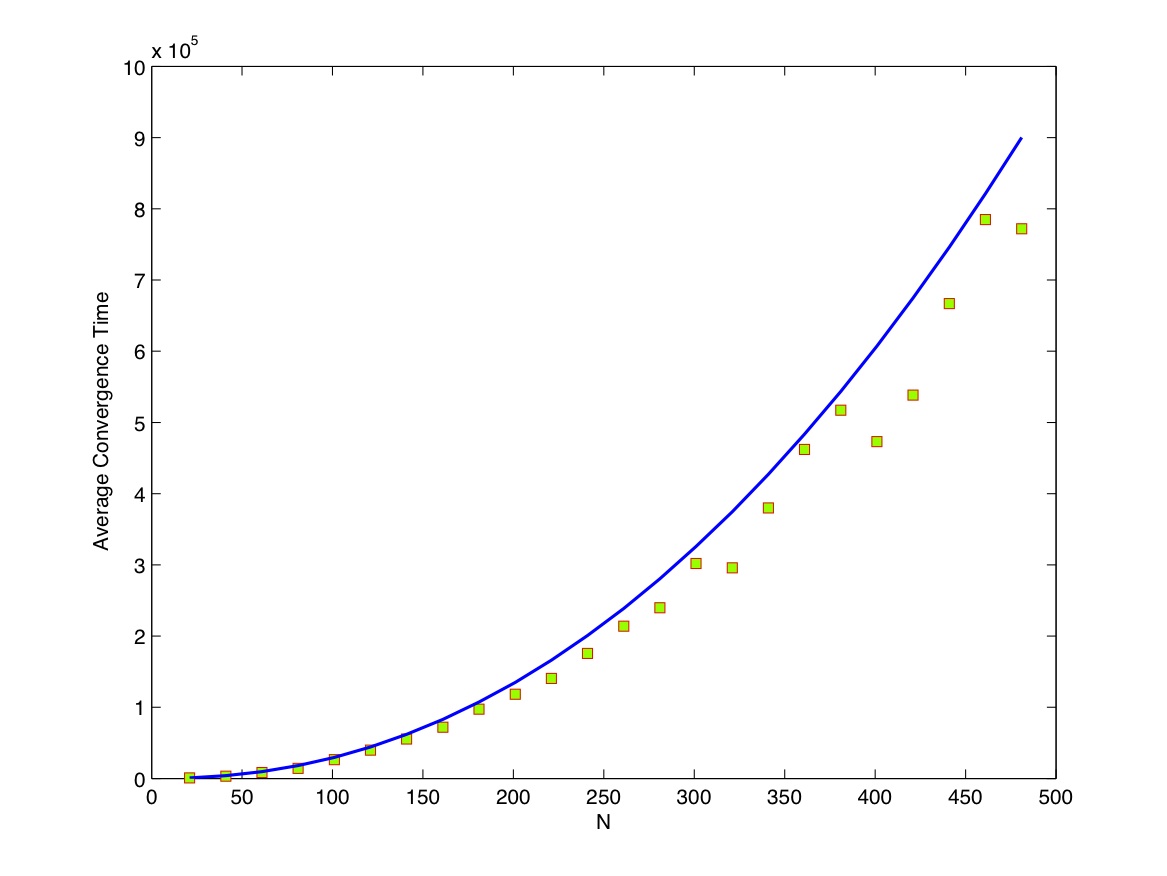}} \hfill
%%%%%%%%%
\subfloat[Simulation results of average convergence time on quantized consensus of star networks in setting 1. The solid line indicates $0.6N^2$, and the dash line indicates $0.7N^2$.]{\label{fig:2}\includegraphics[width=0.32\textwidth]{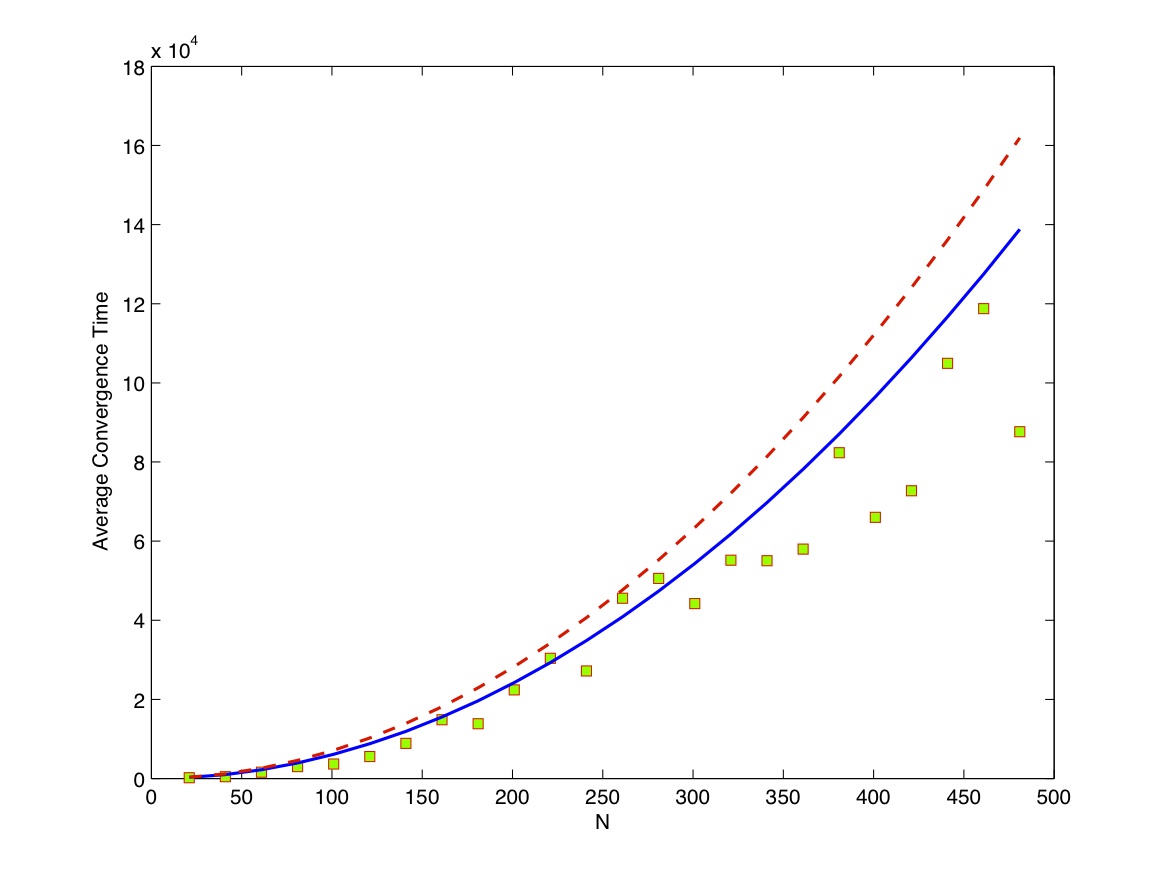}} \hfill
%%%%%%%%%%%
\subfloat[Simulation results of average convergence time on quantized consensus of star networks in setting 2. The solid line indicates $13N\log N$, and the dash line indicates $15N\log N$. ]{\label{fig:3}\includegraphics[width=0.32\textwidth]{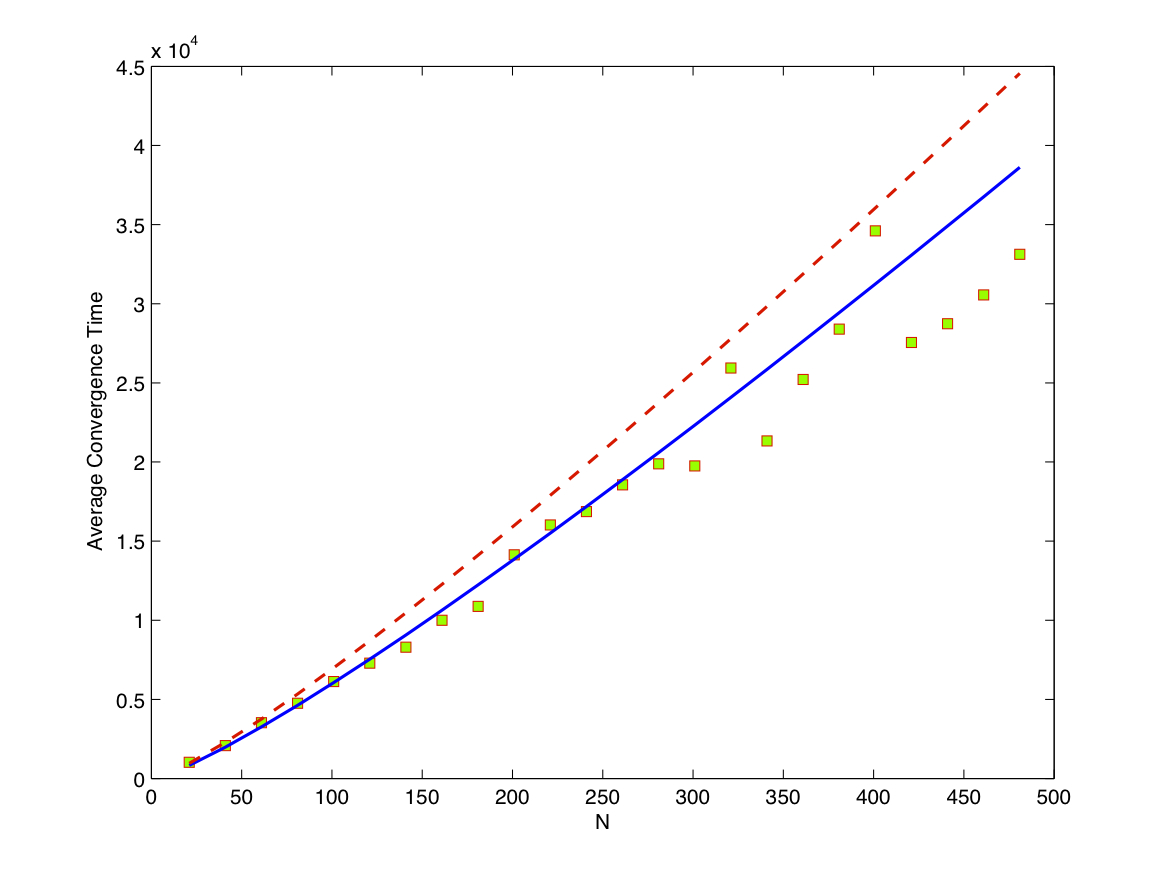}}  \\
%%%%%%%%%%%%
\subfloat[Simulation results of average convergence time on binary consensus algorithm of line graphs. The solid line indicates $0.15N^3\log N$.]{\label{fig:4}\includegraphics[width=0.32\textwidth]{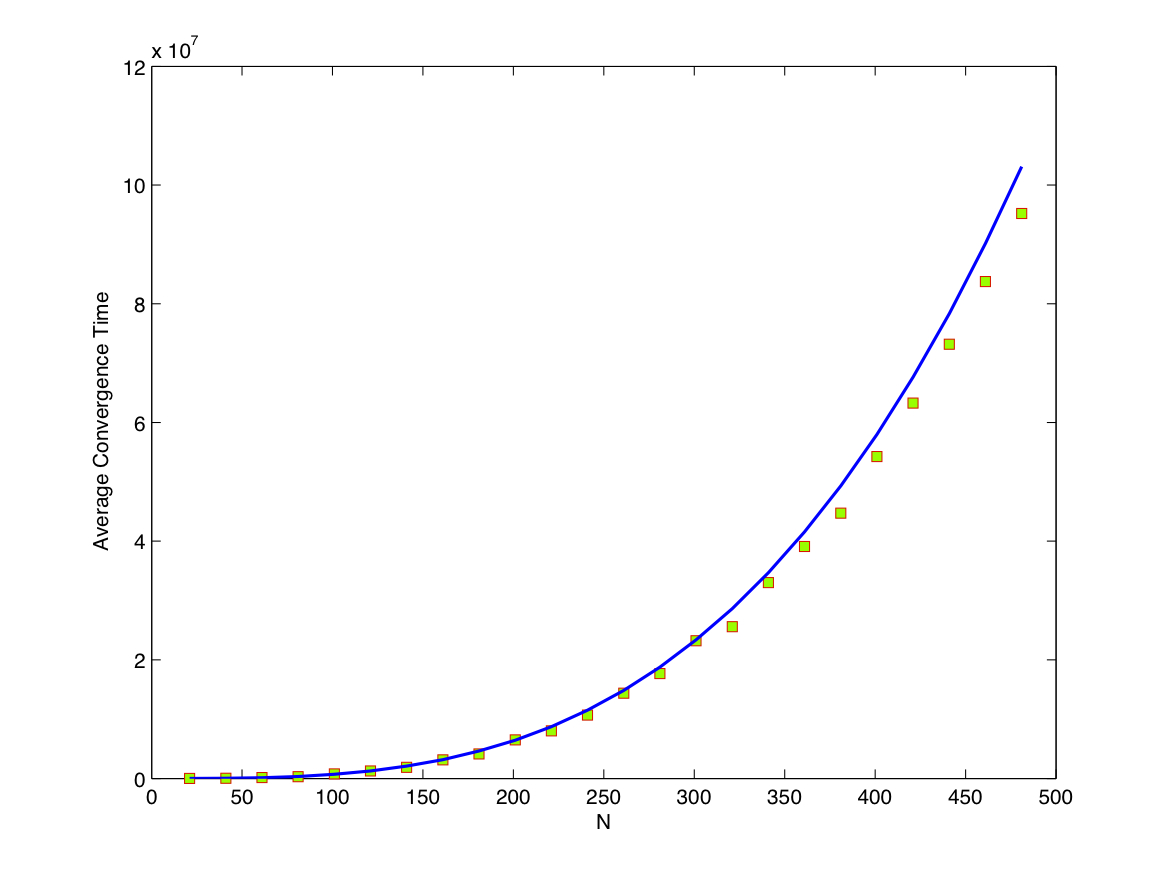}} \hfill
\subfloat[Simulation results of average convergence time on quantized consensus of line graphs in setting 1. The solid line indicates $0.17N^3$.]{\label{fig:5}\includegraphics[width=0.32\textwidth]{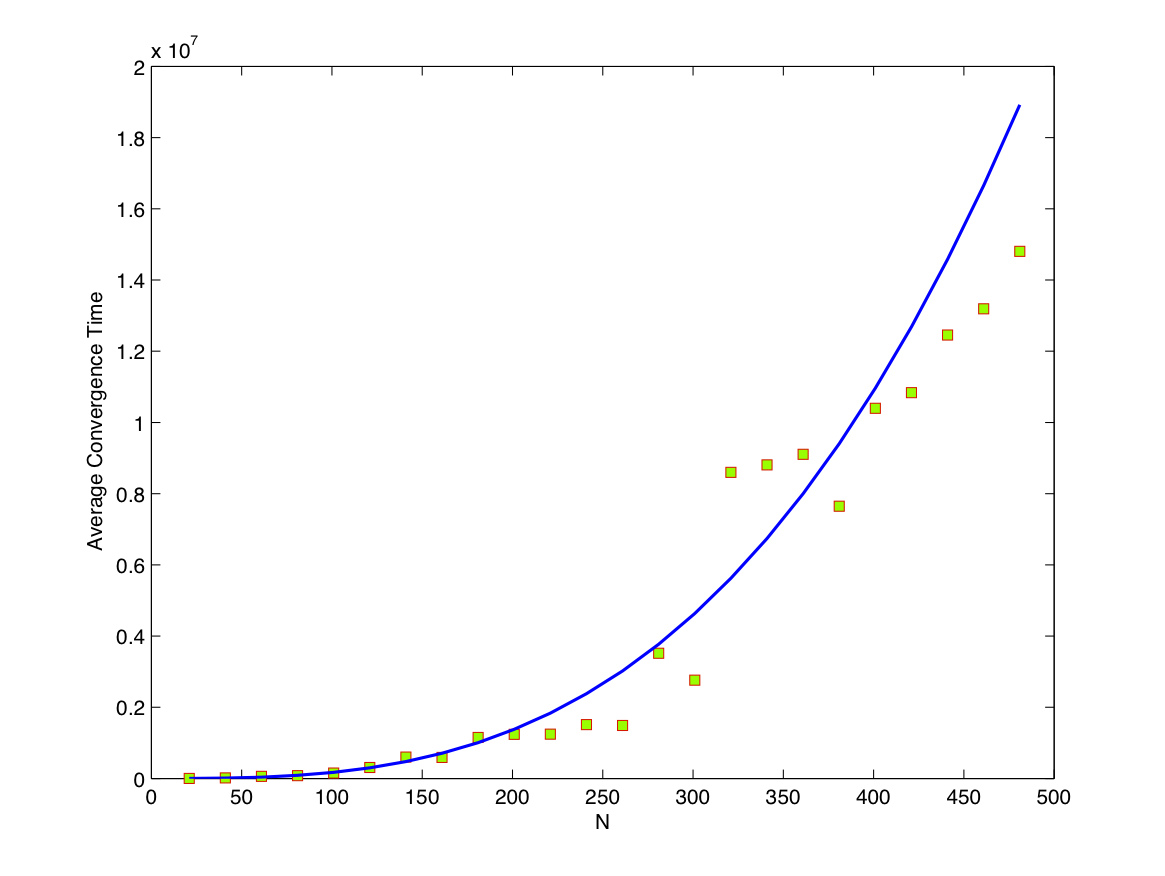}} \hfill
\subfloat[Simulation results of average convergence time on quantized consensus of line graphs in setting 2. The solid line indicates $0.25N^3$.]{\label{fig:6}\includegraphics[width=0.32\textwidth]{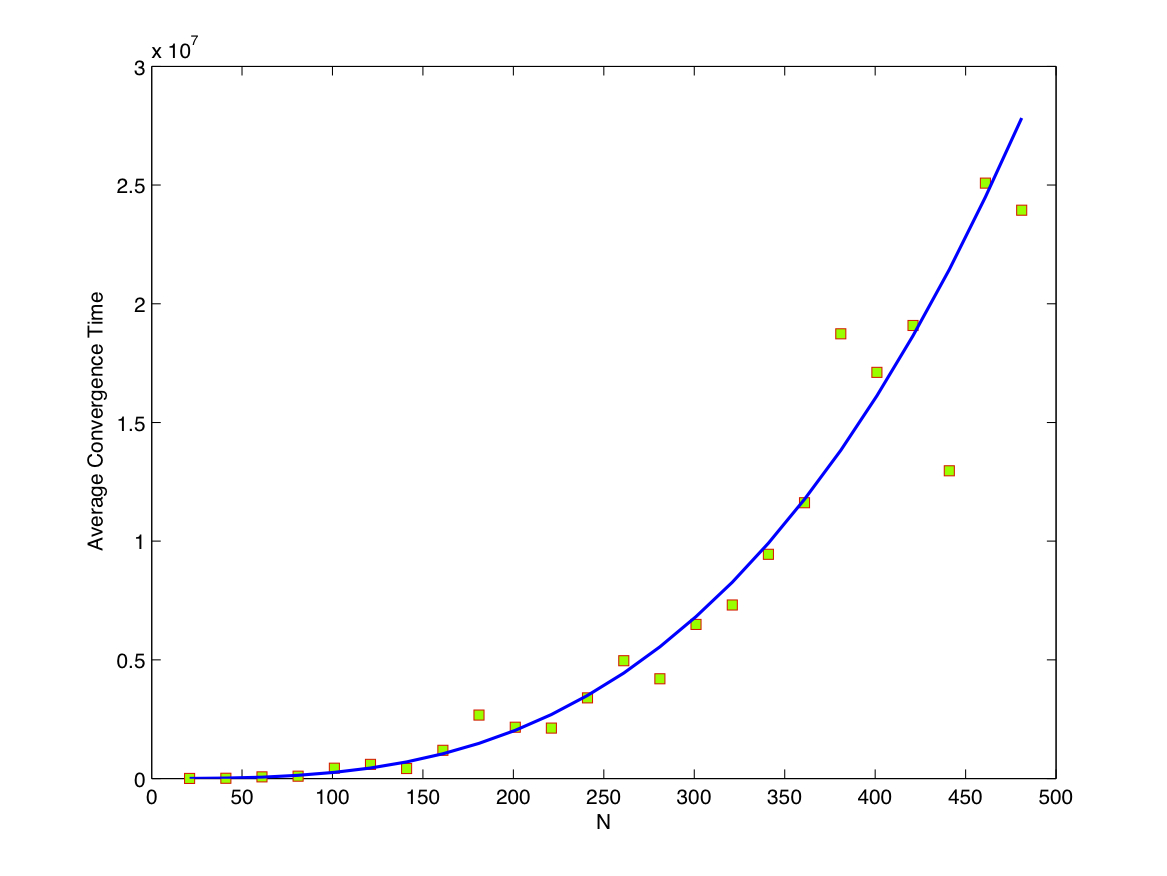}} \\
%%%%%%%%%%%%%%%%%%%%
 \caption{Simulation results on average convergence time (squares) of binary consensus and quantized consensus versus the size of networks.}%The initially active nodes are in State "like" with probability 0.7 and in State "dislike" with probability 0.3. }
 \label{fig:fig}
\end{figure*}

\begin{figure*}[!t]
\ContinuedFloat
\centering
\subfloat[Simulation results of average convergence time on binary consensus algorithm of lollipop graphs. The solid line indicates $0.14N^3\log N$.  ]{\label{fig:7}\includegraphics[width=0.32\textwidth]{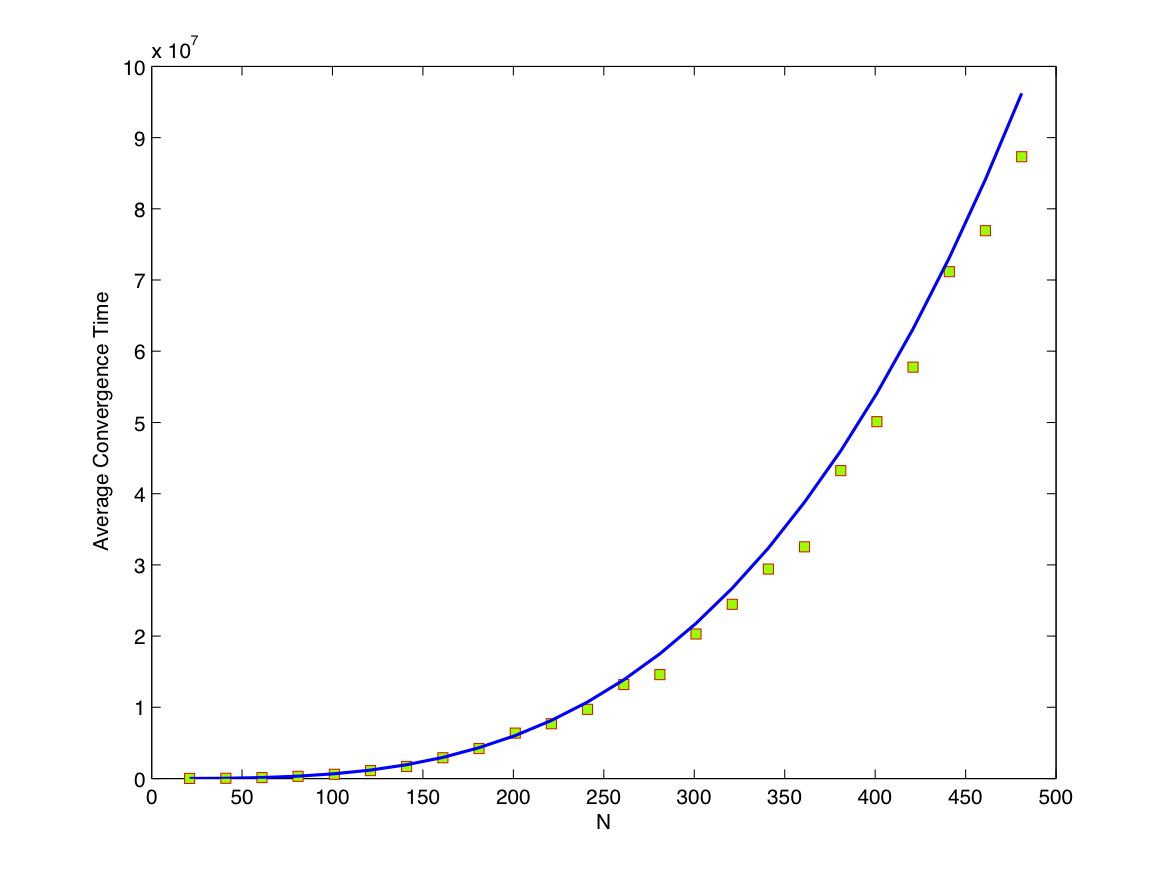}}        \hfill         
\subfloat[Simulation results of average convergence time on quantized consensus of lollipop graphs in setting 1. The solid line indicate $0.15N^3$. ]{\label{fig:8}\includegraphics[width=0.32\textwidth]{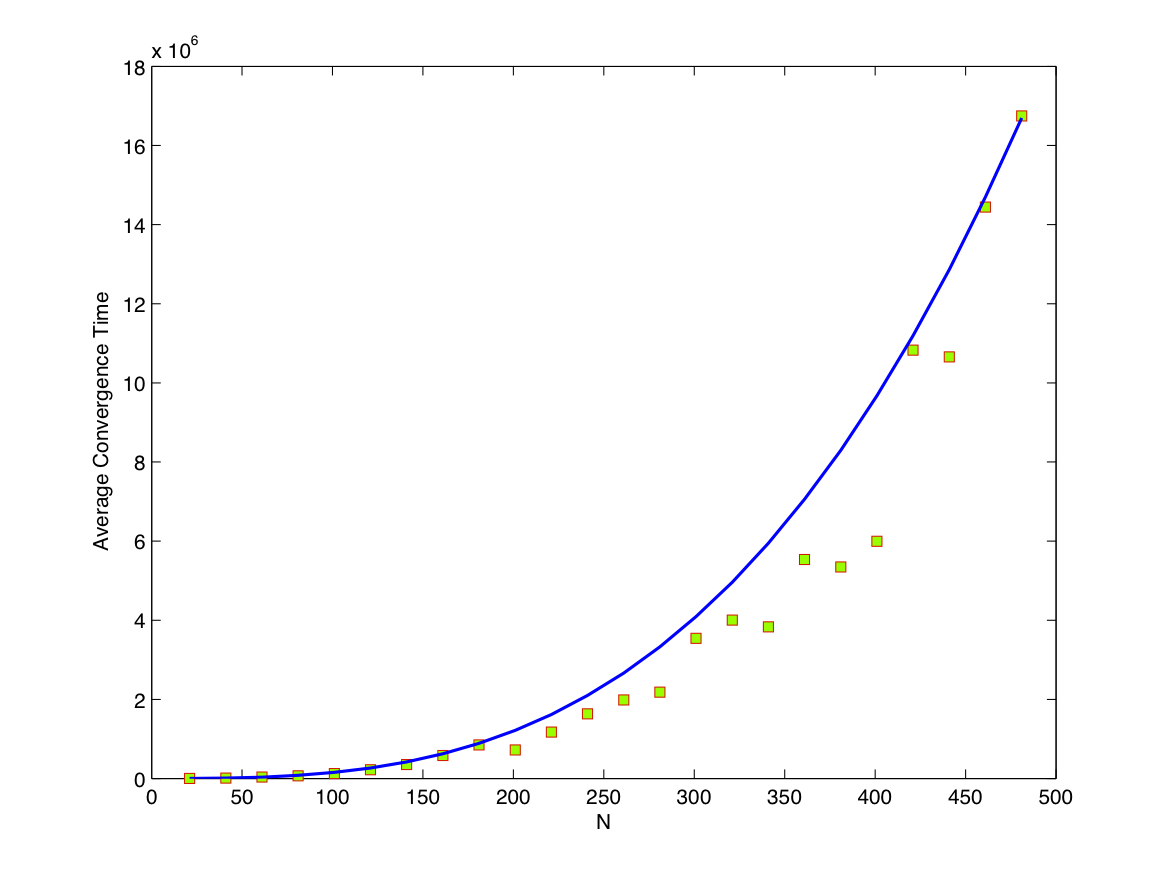}} \hfill
\subfloat[Simulation results of average convergence time on quantized consensus of lollipop graphs in setting 2. The solid line indicate $0.3N^3$. ]{\label{fig:9}\includegraphics[width=0.32\textwidth]{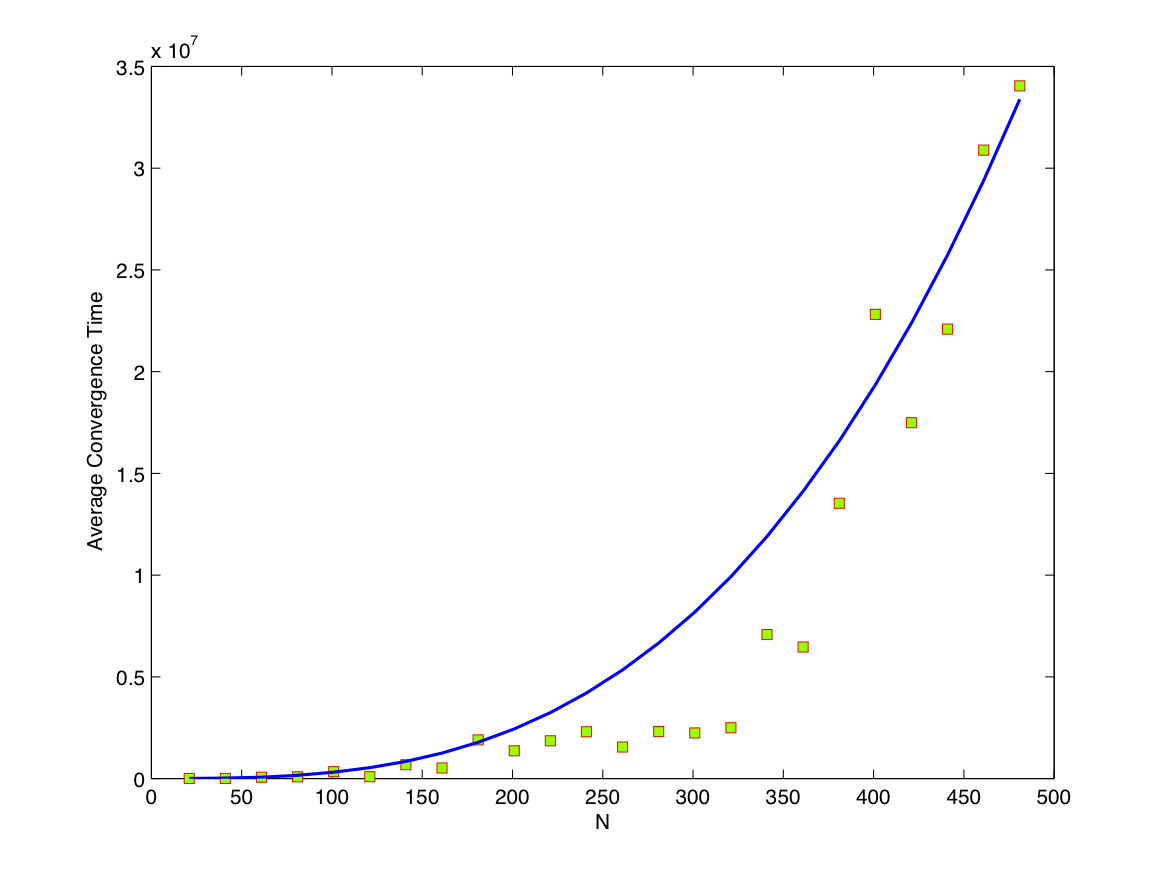}} \\
\subfloat[Simulation results of average convergence time on binary consensus algorithm of Erd\"os-R\'enyi random graphs. The solid line indicates $2N^2\log N$, and the dash line indicates $2.3N^2\log N$. ]{\label{fig:10}\includegraphics[width=0.32\textwidth]{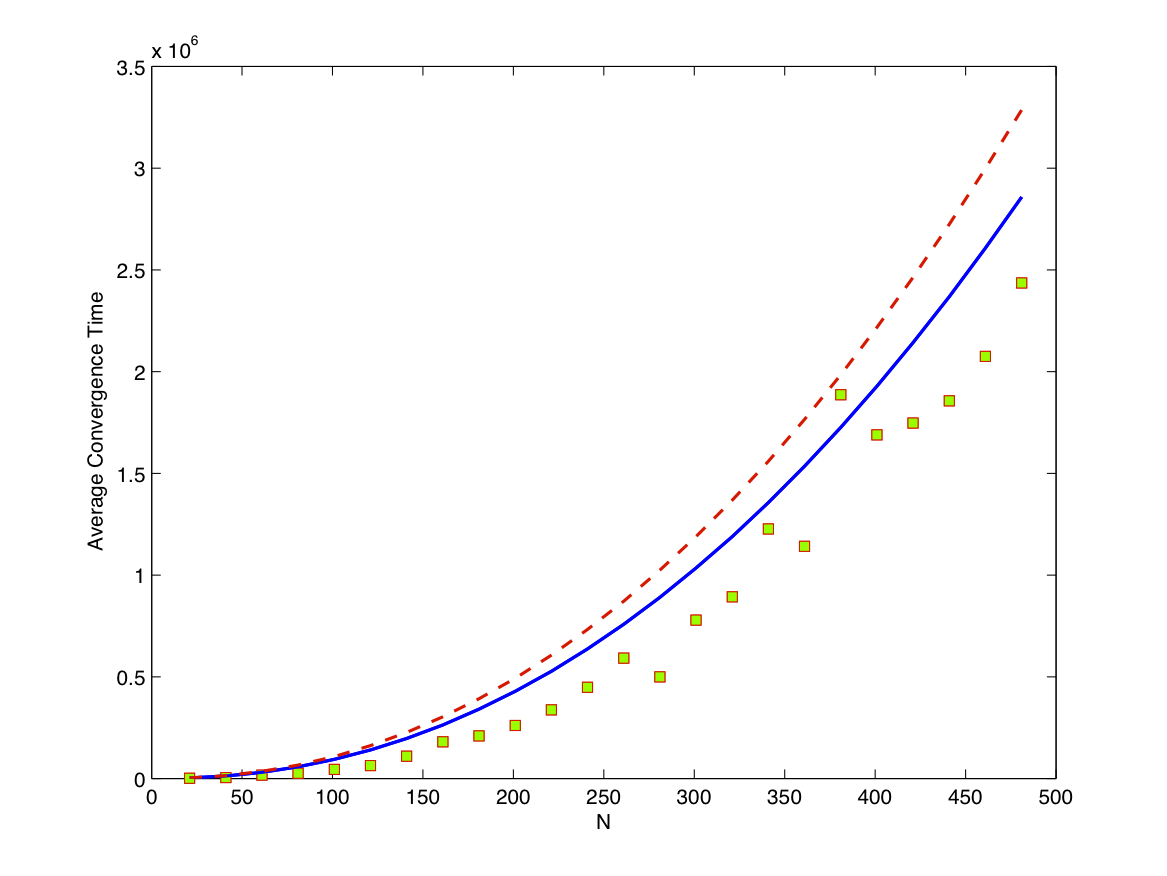}} \hfill
\subfloat[Simulation results of average convergence time on quantized consensus of Erd\"os-R\'enyi random graphs in setting 1. The solid line indicates $0.5N^2$, and the dash line indicates $0.7N^2$. ]{\label{fig:11}\includegraphics[width=0.32\textwidth]{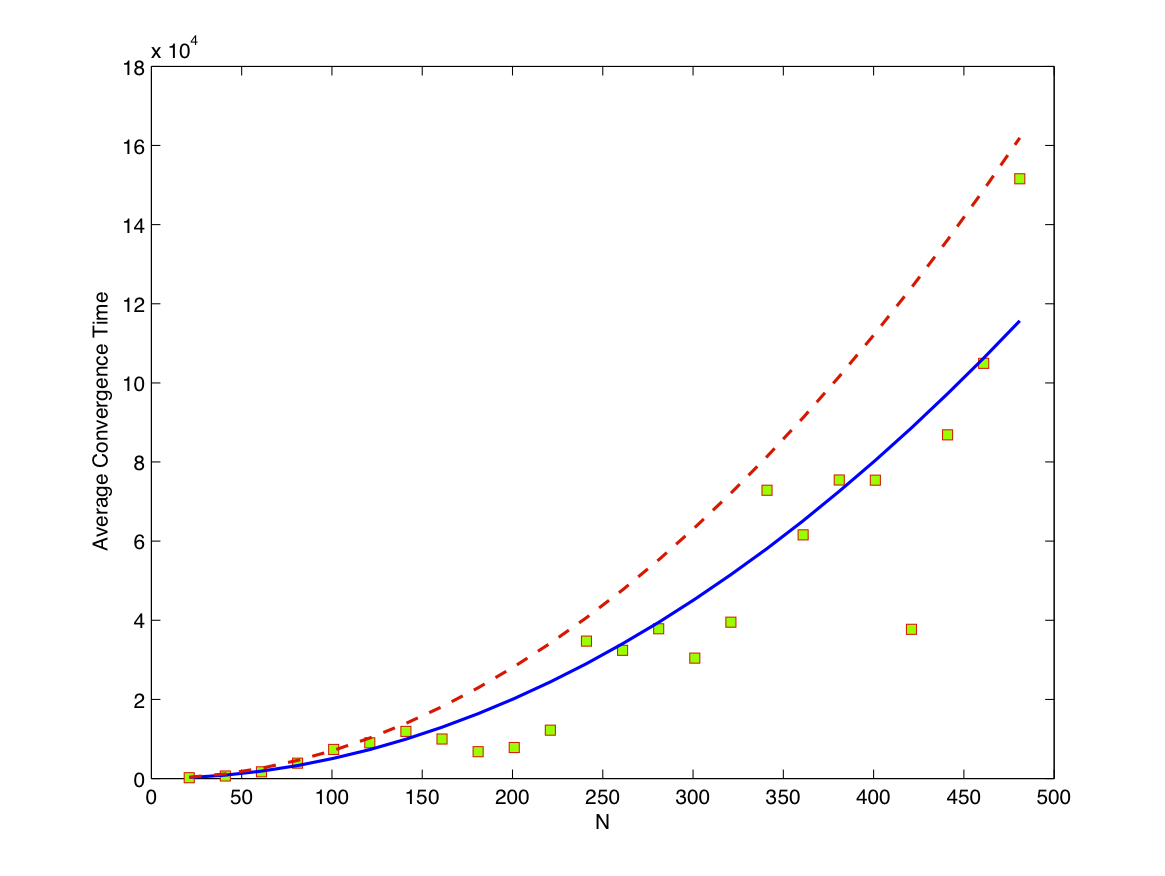}}        \hfill         
\subfloat[Simulation results of average convergence time on quantized consensus of Erd\"os-R\'enyi random graphs in setting 2. The solid line indicates $1.1N^2$. ]{\label{fig:12}\includegraphics[width=0.32\textwidth]{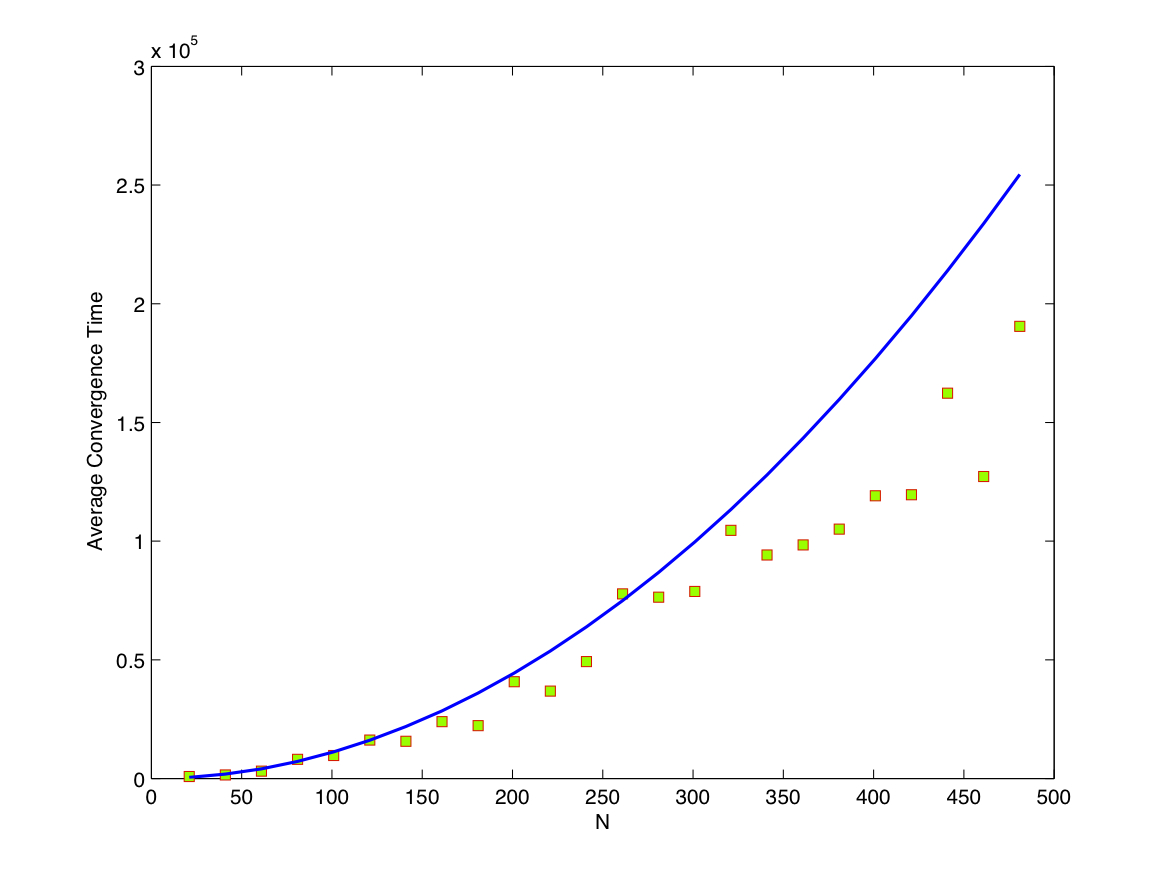}}\\
 \caption{Simulation results on average convergence time (squares) of binary consensus and quantized consensus versus the size of networks (continued).}%The initially active nodes are in State "like" with probability 0.7 and in State "dislike" with probability 0.3. }
 \label{fig:fig}
\end{figure*}

\end{document}